\documentclass[pre,aps,amsmath,twocolumn,showpacs,superscriptaddress,longbibliography]{revtex4-1}

\usepackage{amssymb}
\usepackage{graphicx}
\usepackage[pdftex,bookmarks,colorlinks,breaklinks]{hyperref}
\hypersetup{linkcolor=red,citecolor=blue,filecolor=dullmagenta,urlcolor=blue}
\usepackage{color}
\usepackage{soul}
\definecolor{indiagreen}{rgb}{0.07, 0.53, 0.03}
\usepackage{amsmath}
\usepackage{mathrsfs}
\usepackage{amsfonts}
\usepackage{float}

\usepackage{multirow}
\usepackage{array}
\graphicspath{{./FIGS/}}

\newcommand{\bra}{\left\langle}
\newcommand{\ket}{\right\rangle}

\newcommand{\p}{\mathcal{P}}
\newcommand{\wdt}{\tau_{\text{w}}}
\newcommand{\rmt}{\text{RMT}}

\usepackage{orcidlink}

\begin{document}

\title{Scattering and transport properties of the three classical Wigner-Dyson ensembles at the Anderson transition}

\author{A.~M. Mart\'inez-Arg\"uello\orcidlink{0000-0002-6422-0673}}
\affiliation{Instituto de F\'isica, Benem\'erita Universidad Aut\'onoma de Puebla, Apartado Postal J-48, 72570 Puebla, Pue., Mexico}

\author{M. Carrera-N\'u\~nez\orcidlink{0000-0002-1526-4864}}
\email{moises.carrera.n@gmail.com}
\affiliation{Departamento de Ciencias Naturales y Exactas, Universidad de Guadalajara, Carretera Guadalajara--Ameca Km. 45.5 C.P. 46600. Ameca, Jalisco, Mexico}

\author{J.~A. M\'endez-Berm\'udez\orcidlink{0000-0002-1748-9901}}
\affiliation{Instituto de F\'isica, Benem\'erita Universidad Aut\'onoma de Puebla, Apartado Postal J-48, 72570 Puebla, Pue., Mexico}


\begin{abstract}

An extensive numerical analysis of the scattering and transport properties of the power-law banded random matrix model (PBRM) at criticality in the presence of orthogonal, unitary, and symplectic symmetries is presented. Our results show a good agreement with existing analytical expressions in the metallic regime and with heuristic relations widely used in studies of the PBRM model in the presence of orthogonal and unitary symmetries. Moreover, our results confirm that the multifractal behavior of disordered systems at criticality can be probed by measuring scattering and transport properties, which is of paramount importance from the experimental point of view. Thus, a full picture of the scattering and transport properties of the PBRM model at criticality corresponding to the three classical Wigner-Dyson ensembles is provided.

\end{abstract}


\maketitle


\section{Introduction}

Scattering and transport properties of disordered mesoscopic conductors have been of interest for a long time (see Refs.~\cite{Lee1985,Beenakker1997,MelloBook} and the references therein). Among the diverse phenomena observed in these systems, the disorder-induced localization-delocalization transition of electronic states, known as Anderson or metal-insulator transition (MIT), has received special attention~\cite{Lee1985,Janssen1994,Imada1998,Evers2008,Rodriguez2010,EdwardsBook}. At the MIT, the energy spectra show anomalous behavior while the electronic states present multifractal characteristics and strong amplitude fluctuations. The latter are usually described by an infinite set of critical exponents and represent one of the most important characteristics of the MIT~\cite{Janssen1994,Evers2008,Janssen1998,Huckestein1995,Mirlin2000a,Fyodorov1994,Fyodorov1995,Falko1995,Wegner1980,Schreiber1991,Mildenberger2002,Backer2019,Carnio2019}. 
At criticality, i.e.~at the MIT, both the dimensionality and the symmetries present in the system play an important role.

The statistical properties of ordinary disordered samples are well described by the random matrix theory (RMT)~\cite{Guhr1998} for which three universal symmetry classes are known: the orthogonal [the symplectic] class describing systems in the presence of time reversal and presence [absence] of spin-rotation symmetry, and the unitary class for systems with broken time-reversal symmetry. In the Dyson scheme these symmetries are labeled by the indices $\beta = 1, 2,$ and 4, for the orthogonal, the unitary, and the symplectic classes; respectively~\cite{MehtaBook,Dyson1962a,Dyson1962b}.

Until now, many important features of disordered systems at the MIT have been analyzed using numerical techniques. This is due to the complexity in obtaining analytical expressions at criticality, some of which are available only perturbatively. In particular, the so-called power-law banded random matrix (PBRM) model has widely been used since it captures all the key features of the Anderson critical point and is also convenient for its low computational  cost~\cite{Evers2008,Mirlin1996,Kravtsov1999,Varga2000}. For the closed system on the one hand, since the appearance of the PBRM model originally proposed by Mirlin \textit{et al}.~\cite{Mirlin1996}, a plethora of studies regarding different aspects of the model with $\beta=1$ symmetry have been performed~\cite{Mirlin1996,Evers2008,Mirlin2000a,Varga2000,Mendez2012,Cuevas2001,Kravtsov1997,Kravtsov2000,Varga2002,Mendez2014,Mirlin2000,Evers2000,Rao2022} while less studies regarding $\beta=2$ symmetry are reported~\cite{Mendez2014,Kravtsov2011,Mirlin2000a,Mirlin2000}. Moreover, the energy spectra and multifractal behavior of the PBRM model in the presence of the symplectic symmetry ($\beta=4$) have recently been analyzed~\cite{Carrera2021}. For the open system, on the other hand, several scattering and transport properties of the model when the system is in the presence of time-reversal invariance ($\beta=1$)~\cite{Antonio2009,Mendez2005,Mendez2006,Mendez2010,Kottos2002,Mendez2005,Mendez2014}, which are in agreement with the ones obtained by using the three-dimensional Anderson model at MIT~\cite{Mendez2010,Cuevas2001}, have been investigated. However, to our knowledge, the analysis of the scattering and transport properties of the PBRM model for the unitary case are scarce~\cite{Alcazar2009} while the symplectic case has been left unexplored.

It is the purpose of the present paper to deepening the understanding of the scattering and transport properties of critical systems belonging to the symplectic class; that is, we study the open symplectic PBRM ensemble at criticality. The regime of a small number of single-mode leads attached to the PBRM model is studied in detail. Nevertheless, the multichannel or multiple single-mode leads setup is also studied for some scattering and transport quantities of interest. In order to provide a full picture of the PBRM model for the three classical Wigner-Dyson ensembles, the scattering and transport properties of the PBRM model for the $\beta=1$ and $\beta=2$ cases, previously considered in the literature, are also reviewed and extended when appropriate. Our results are also compared with RMT predictions in the appropriate limits.

The organization of the paper is as follows. In the next section, the generalized PBRM model in the presence of the three symmetry classes and the scattering setup  are described. The perfect coupling regime, the Wigner delay time, and the resonance widths when the PBRM supports one open channel are analyzed in detail in Sec.~\ref{sec:onesinglechannel}. This is of particular interest, since it shows that the multifractal properties of the isolated PBRM model can be probed by measuring transport properties. The analysis of the scattering and transport properties of the PBRM model in the two and four open-channel setups is the subject of Sec.~\ref{sec:MChannels}. In the same section, the scattering and transport properties in the multichannel setup are also presented. Finally, the conclusions are given in Sec.~\ref{sec:Conclusions}.


\section{Model and scattering setup}
\label{sec:Model}

The PBRM model describes one-dimensional (1D) tight-binding wires of length $N$ with random long-range hoppings \cite{Evers2008,Mirlin1996}. In the presence of the three symmetry classes, it is represented by $N\times N$ real symmetric ($\beta=1$), complex Hermitian ($\beta=2$), or $2N\times 2N$ quaternion-real Hermitian ($\beta=4$) matrices whose elements are statistically independent random variables drawn from a normal distribution with zero mean and variance given by~\cite{Carrera2021}
\begin{align}
\langle |H_{ii}|^{2} \rangle & =  \beta^{-1} \quad \mathrm{and} \nonumber \\
\langle |H_{ij}|^2\rangle & =  \frac{1}{2(1 + \delta_{\beta, 4})}\frac{1}{ 1+
\left[ \sin\left(\frac{\pi|i-j|}{N}\right)\Big/
\left(\frac{\pi b}{N}\right) \right]^{2\mu}} .
\label{eq:Hij}
\end{align}
 The matrix sizes are $L=N$ for $\beta = 1$ and 2, and $L=2N$ for $\beta=4$. The PBRM model is a random matrix ensemble with off-diagonal matrix elements decaying away from the diagonal in a power-law fashion. Also, in (\ref{eq:Hij}) the PBRM model is in its periodic version, i.e., the 1D wire is in a ring geometry, where $\mu$ and $b$ are the  parameters of the model and $\delta_{\beta, 4}$ is the Kronecker delta. In particular, for the symplectic case ($\beta=4$) the PBRM model preserves the quaternion structure of the Hamiltonian where each eigenvalue is two-fold degenerate due to Kramers degeneracy (for more details see Ref.~\cite{Carrera2021}). The power-law decay $\mu = 1$ sets the PBRM model at the MIT critical point~\cite{Evers2008,Mirlin2000a,Mirlin1996,Varga2000,Cuevas2001,Carrera2021,Kravtsov1997,Kravtsov2000,Varga2002}. Furthermore, regardless of the value of $\mu$, insulating- to metallic-like behavior may be induced by varying the effective bandwidth $b$ from small $b$ to large $b$ values, respectively. Here, the scattering and transport properties of the PBRM model of Eq.~(\ref{eq:Hij}) at criticality, $\mu=1$, is the focus of this work.

The isolated wire is opened by attaching to it $M$ semi-infinite single-channel leads, each one described by a 1D tight-binding Hamiltonian
\begin{equation}
\label{eq:Hamlead}
H_\text{lead}=\sum_{n=1}^{-\infty}(|n\rangle\langle n+1|+|n+1\rangle\langle n|),
\end{equation}
thus $M$ establishes the number of open channels or propagating modes. The $M\times M$ scattering matrix, $S(E)$, can be written as~\cite{Mahaux1969,Verbaarschot1985,Rotter1991,Kottos2002}
\begin{equation}
S(E) =\left(
\begin{array}{ccc}
r & t' \\
t & r'
\end{array} 
\right) \\
= \openone_{M} - 2i\pi 
W^{T}(E \openone_{L} - H_\text{eff})^{-1}W,
\label{eq:scatteringmatrix2}
\end{equation}
where $r$ [$r'$] and $t$ [$t'$] are the reflection and transmission amplitudes when the incidence is from the left [right], $\openone_{n}$ stands for the unit matrix of dimension $n$, $E$ is the energy, and the superscript $T$ indicates the matrix transposition operation. In Eq.~(\ref{eq:scatteringmatrix2}), $H_\text{eff}$ is the non-Hermitian effective Hamiltonian given by
\begin{equation}
\label{eq:Heff}
H_\text{eff} = H-i\pi WW^{T} ,
\end{equation}
where $H$ is the $L\times L$ Hamiltonian matrix of the PBRM model that describes the isolated wire with $L$ resonant states and $W$ is an $L\times M$ energy independent matrix that couples those resonant states to the $M$ propagating modes in the leads. The elements of $W$ are $W_{ij} =w_{0}\delta_{ij_{0}}$, where $w_{0}$ is the coupling strength between the wire and the leads and $j_{0}=1,\ldots,M$  are the sites at which the leads are attached. According to the symmetry present in the Hamiltonian, the $S$ matrix is unitary symmetric, unitary, and unitary self-dual matrix for $\beta = 1, 2, $ and 4, respectively~\cite{Beenakker1997}. Additionally, due to the ring geometry of the isolated wire under consideration, the scattering and transport properties do not depend on which site the leads are attached to. Then, for simplicity and without loss of generality, in this work the leads are attached at consecutive sites of the wire.


\section{PBRM model with one open channel}
\label{sec:onesinglechannel}

In this section, the statistical properties of scattering phases, Wigner delay time, and resonance widths, when the scattering system supports one open channel and is in the presence of the $\beta=1, 2,$ and 4 symmetries are analyzed. The relation between Wigner delay times and the properties of the spectra and eigenstates of the corresponding isolated wire are also discussed. 


\subsection{Perfect coupling regime}

%
\begin{figure}
\centering
\includegraphics[width=0.48\textwidth]{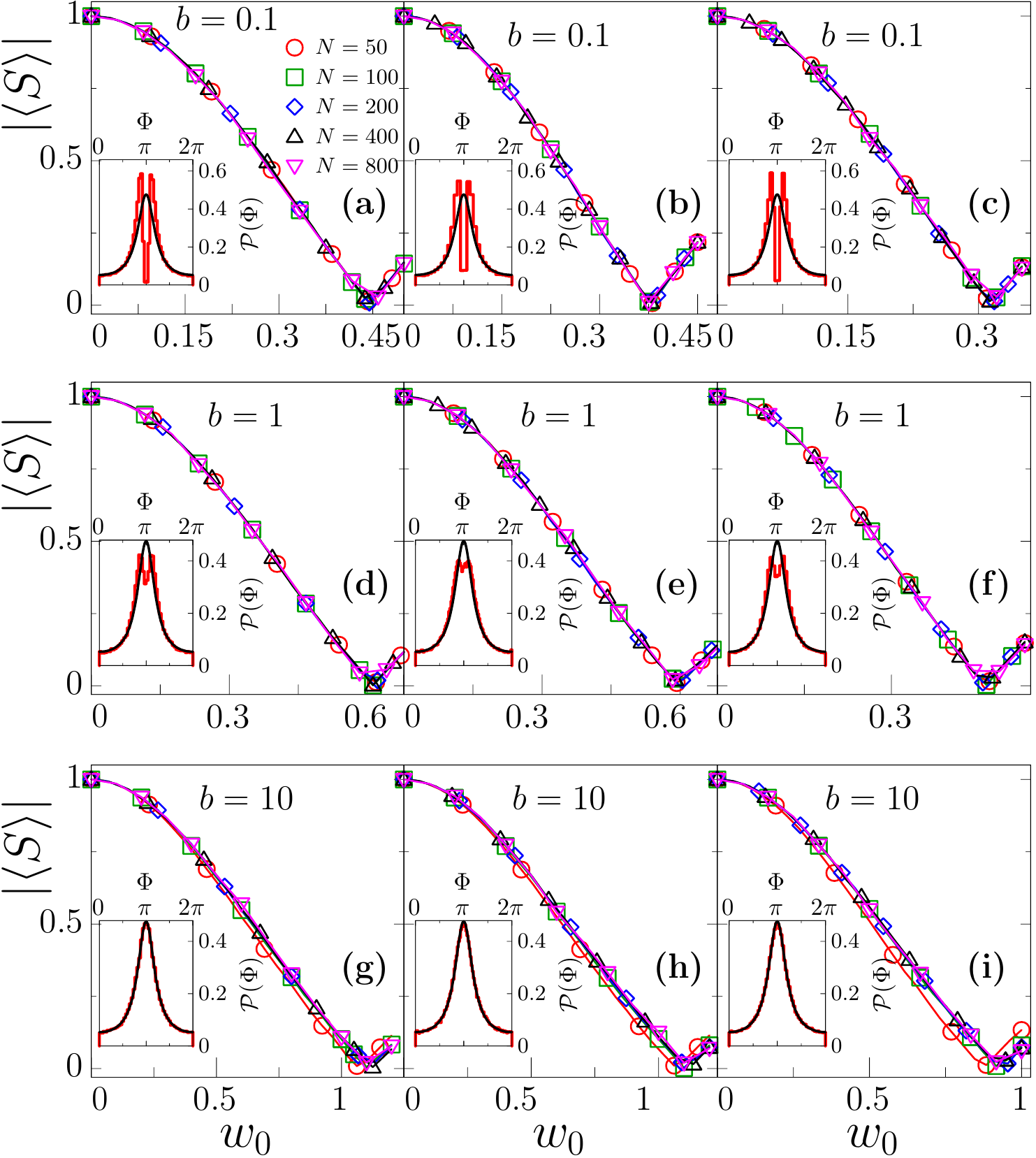}
\caption{Absolute value of the optical $S$ matrix for the PBRM model at criticality as a function of the coupling strength, $w_{0}$, for values of $b$ indicated in each panel and several wire lengths $N$ as indicated in panel (a). The symmetries are $\beta=1$ (first column), 2 (second column), and 4 (third column). The insets show numerically obtained histograms (red lines) compared with the phase distribution~(\ref{eq:PDFofPhi}) (black continuous lines) for $|\langle S\rangle| = 0.5$. The error bars are smaller than the symbol size, so they are not displayed.}
\label{fig:Savg}
\end{figure}

In the one-channel setup, the scattering matrix of Eq.~(\ref{eq:scatteringmatrix2}) reduces to a phase given by $S(E) = \mathrm{e}^{\mathrm{i} \Phi(E)}$. This case corresponds to a single-channel lead attached to the wire~(\ref{eq:Hij}). For the PBRM model in the limit $b\gg1$ (metallic-like regime), that phase is distributed according to the following expression~\cite{Ossipov2005}
\begin{equation}
\label{eq:PDFofPhi}
\p(\Phi) = \frac{1}{2\pi} \frac{1}{\gamma+\sqrt{\gamma^{2}-1}\cos\Phi},
\end{equation}
where $\gamma=(1+\vert\bra S\ket\vert^{2})/(1-\vert\bra S\ket\vert^{2})$. In the so-called \emph{perfect coupling regime}, the averaged scattering matrix $\bra S\ket$, also known as optical matrix, vanishes and the phase is uniformly distributed over the unit circle.

Before proceeding with the analysis of the phase distribution~(\ref{eq:PDFofPhi}), some general statements about the numerical simulations are given. For the statistical analysis shown throughout this work, most calculations are performed in the perfect coupling regime since in this limit a number of analytical predictions from RMT are known. Also, the calculations are performed around $E \sim 0$ considering wire lengths of $N = 50, 100, 200, 400$, and 800 with $10^{6}, 10^{6}, 10^{5}, 10^{5}$, and $10^{5}$ random realizations, respectively. All fittings are performed through the nonlinear least-squares Marquardt-Levenberg algorithm as implemented by gnuplot. The error bars are computed by the jackknife method, unless specified otherwise. Histograms for the probability distributions and for the cumulative probabilities do not contain error bars since the statistics is done with a large amount of data such that the error is not significant.

%
\begin{figure}
\centering
\includegraphics[width=0.48\textwidth]{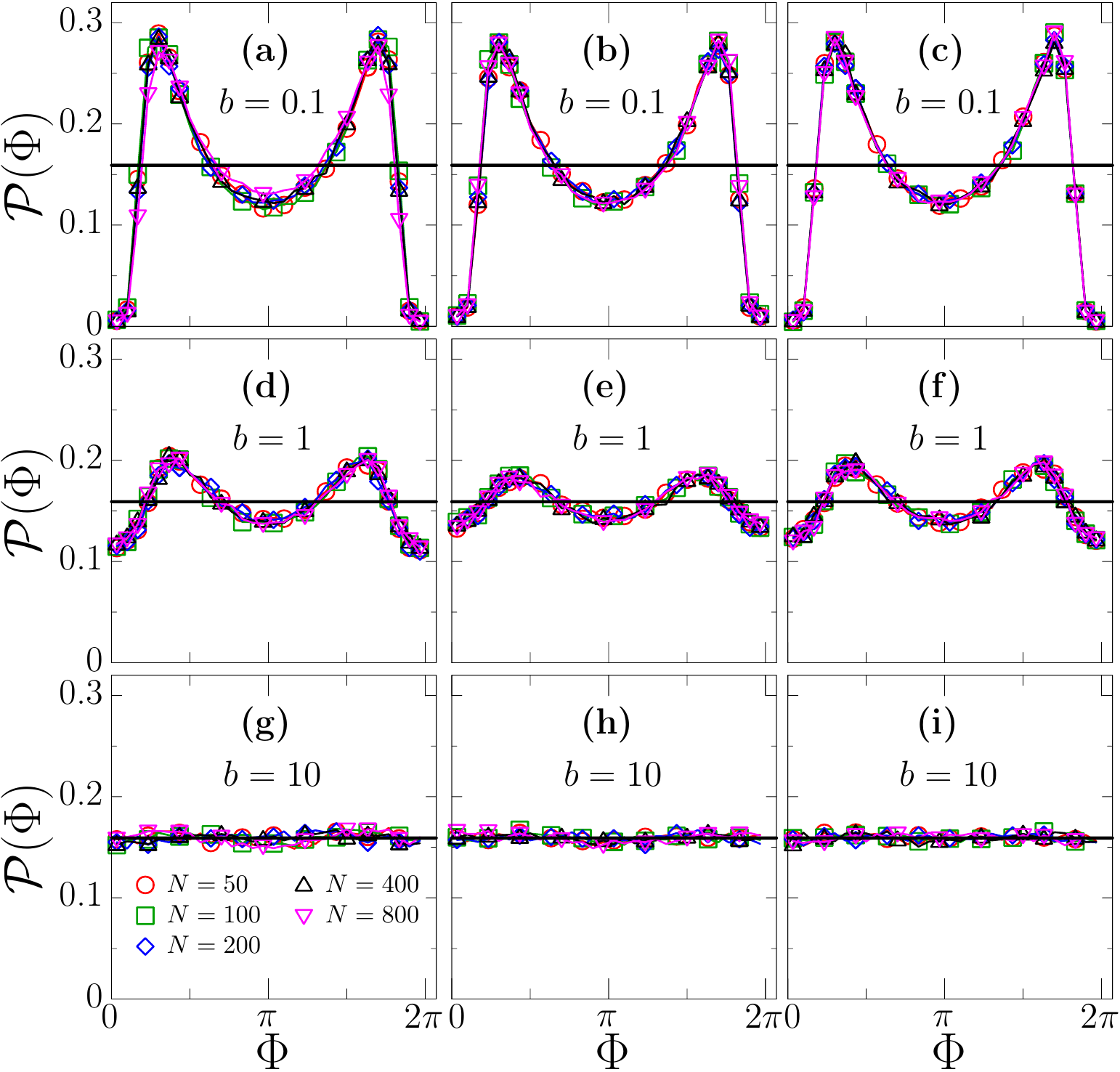}
\caption{Phase distribution in the perfect coupling regime, $|\langle S\rangle|~\sim0$, for the PBRM model at criticality for several wire lengths $N$ as indicated in panel (g) and several values of $b$, as indicated in each panel. The symmetries are $\beta=1$ (first column), 2 (second column), and 4 (third column). Symbols correspond to numerical results while  black lines correspond to the RMT prediction~(\ref{eq:PDFofPhi}).}
\label{fig:Phases}
\end{figure}

In Fig.~\ref{fig:Savg}, the modulus of the optical matrix as a function of the coupling strength $w_{0}$ for the PBRM model at criticality with $\beta=1$ (first column), $\beta=2$ (second column), and $\beta=4$ (third column) is shown. Different values of the bandwidth $b$ and wire lengths $N$ are considered. It is observed that the perfect coupling, $|\langle S\rangle|\approx0$, does not depend on the wire length, but shows a strong dependence on $b$. The perfect coupling is attained for values of $w_0 < 0.5$, $0.45 < w_0 < 1$, and $w_0\approx1$, when the system is in the insulator-like (top panels), in between the insulator-like and the metallic-like (middle panels), and close to the metallic-like regime (bottom panels), that is for $b=0.1$, 1, and 10, respectively. In the insets, the phase distribution $\p(\Phi)$ is shown for a coupling strength $|\langle S(w_{0})\rangle| \approx 0.5$. The histograms in red lines correspond to numerical results while the continuous black lines correspond to the analytical distribution of Eq.~(\ref{eq:PDFofPhi}). For $b=0.1$ and 1, the histograms show two peaks around $\Phi=\pi$ which vanish for $b=10$, i.e., when the system displays a metallic behavior. In the latter, a good agreement with the RMT prediction is observed.

The phase distribution in the perfect coupling regime for the PBRM model at criticality with $\beta=1$ (first column), 2 (second column), and 4 (third column) symmetries is shown in Fig.~\ref{fig:Phases}. The wire lengths are indicated in 
Fig.~\ref{fig:Phases}(g) while the values of parameter $b$ are indicated in each panel. The symbols are obtained by numerical simulations while the solid lines are the RMT prediction of Eq.~(\ref{eq:PDFofPhi}). Again, two big peaks around $\Phi=\pi$ show up when the system is close to the insulator-like regime $b=0.1$ (top panels) which tend to disappear as $b$ increases (middle and bottom panels). When the system attains a metallic behavior, $b=10$ (bottom panels), the phase is uniformly distributed around $1/2\pi$, in accordance with the RMT prediction~(\ref{eq:PDFofPhi}). It is also clear that $\p(\Phi)$ does not depend on the wire length nor the symmetry (orthogonal, unitary, or symplectic), as expected in the one open-channel setup~\cite{MelloBook}.

%
\begin{figure}
\centering
\includegraphics[width=0.48\textwidth]{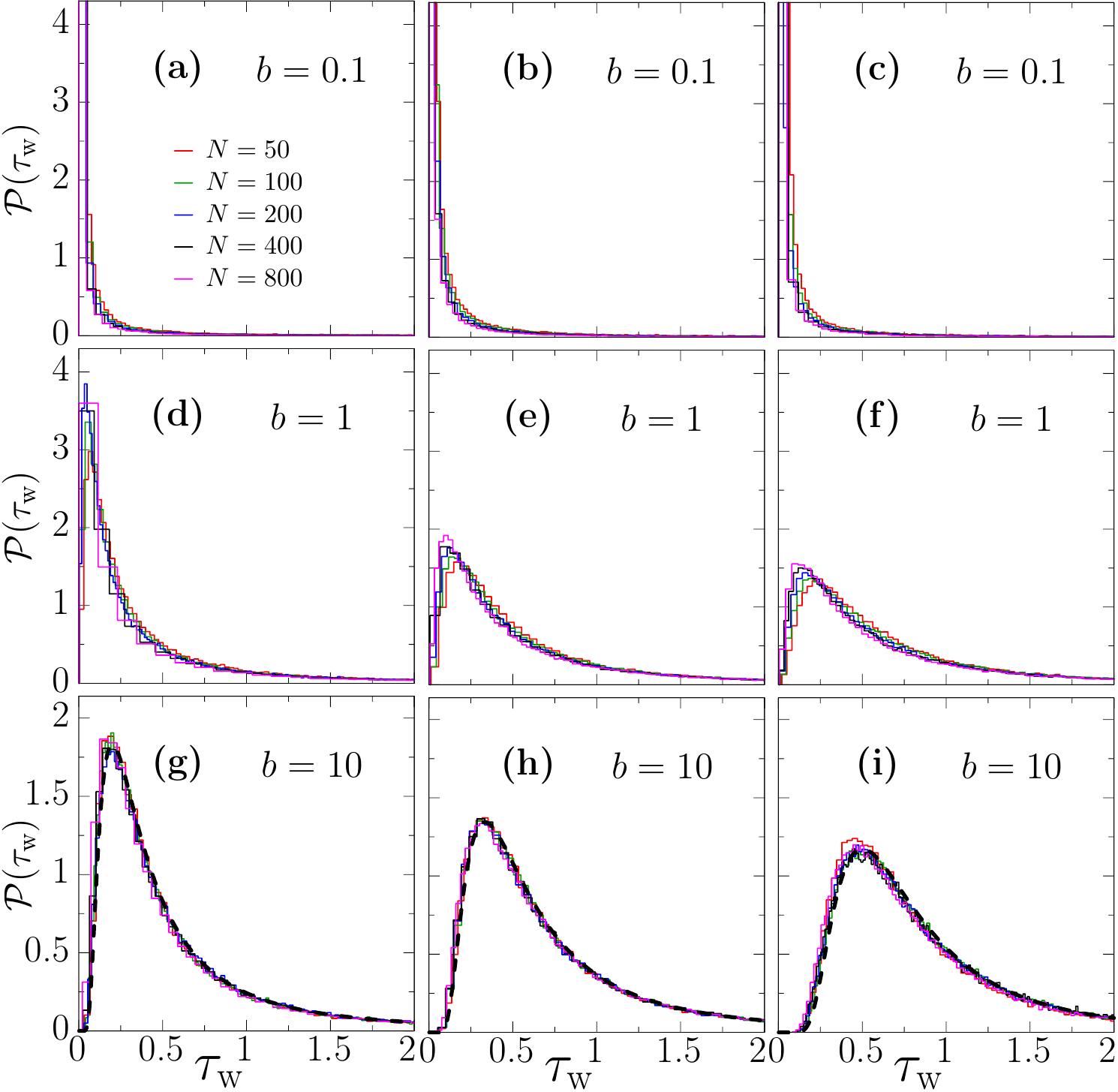}
\caption{Distribution of the Wigner delay time for the PBRM model at criticality for several wire lengths $N$, as indicated in panel (a). The symmetries are $\beta=1, 2$, and 4, in the first, second, and third columns, respectively. The different values of $b$ are indicated in each panel. The histograms are obtained through Eq.~(\ref{eq:Wdtch}), with the normalization constant  $w_{0}^{2}\pi/(2L)$, in the perfect coupling regime. The dashed black lines (bottom panels) correspond to the RMT prediction of Eq.~(\ref{eq:WSdt1ch}).}
\label{fig:Wdt-1ch}
\end{figure}
%


\subsection{Wigner delay time and resonance widths}

The delay experienced by a quantum particle due to its interactions with a scattering region is described by the so-called Wigner delay time $\wdt$. Near the center of the spectrum ($E=0$), it is given by~\cite{Steinbach2000,Kottos2002}
\begin{equation}
\wdt(E=0) = \left.\frac{d\Phi(E)}{dE}\right\vert_{E=0} = -2\;\text{Im}\;\text{Tr} \left.\left( E-H_\text{eff} \right)^{-1}\right\vert_{E=0}.
\label{eq:Wdtch}
\end{equation}

In the metallic regime and for the one open-channel setup, the distribution of the Wigner delay time is known for all symmetry classes $\beta=1, 2,$ and 4~\cite{Gopar1996,Fyodorov1997a,Fyodorov1997b,Angel2017}, namely
\begin{equation}
\label{eq:WSdt1ch}
\p(\wdt) = \frac{2/\beta}{(\beta /2)!} 
\left( \frac{\beta}{2\wdt} \right)^{2 + \beta/2} \mathrm{e}^{-\beta/2\wdt} .
\end{equation}

On the other hand, the poles of the scattering matrix show up as resonances which in turn are the complex eigenvalues $\mathcal{E}_{n}=E_{n} -i\Gamma_{n}/2$ of the effective non-Hermitian Hamiltonian $H_\text{eff}$ [see Eq.~(\ref{eq:scatteringmatrix2})], with $E_{n}$ and $\Gamma_{n}$ respectively the position and width of the $n$th resonance. Furthermore, the resonance width $\Gamma_{n}$ is related to the life time of the $n$th resonance as $\tau_{n}=1/\Gamma_{n}$, and hence a relationship between $\Gamma_{n}$ and the Wigner delay time is expected. These quantities, delay times and resonance widths, have been of pivotal importance in the realm of complex scattering both theoretically and experimentally~\cite{Mendez2005,Kottos2002,Steinbach2000,Gopar1996,Fyodorov1997a,Fyodorov1997b,Angel2017,Kukulin1989,Doron1990,Kottos2005,Fyodorov2012,Kuhl2008,Fyodorov2015,Novaes2022}, where recent progress has been made to extend their study to wave-chaotic scattering systems in the presence of absorption~\cite{Chen2021a,Chen2021b}. For the PBRM model at the critical point, to our knowledge, exact theoretical results for the resonance statistics are scarce, nonetheless recent progress has been reported for resonance statistics in standard banded matrices in both weak and strong localization regimes~\cite{Fyodorov2022}.

The distribution of the Wigner delay time for the PBRM model at criticality is reported in Fig.~\ref{fig:Wdt-1ch} for the three symmetry classes $\beta=1, 2$, and 4, in the first, second, and third column, respectively. The values of the bandwidth $b$ are shown in each panel. For $b=0.1$ (top panels), $\p(\wdt)$ has its maximum at $\wdt \sim 0$; that is, the system is in the localized regime and conduction is suppressed. For $b=1$ (middle panels), relatively small time delays dominate, meaning that the system is neither an insulator nor a conductor. For $b=10$ (bottom panels), the distribution of $\wdt$ is well described by its RMT prediction~(\ref{eq:WSdt1ch}) which sets the system in a metallic-like regime.

\begin{figure}
\centering
\includegraphics[width=0.48\textwidth]{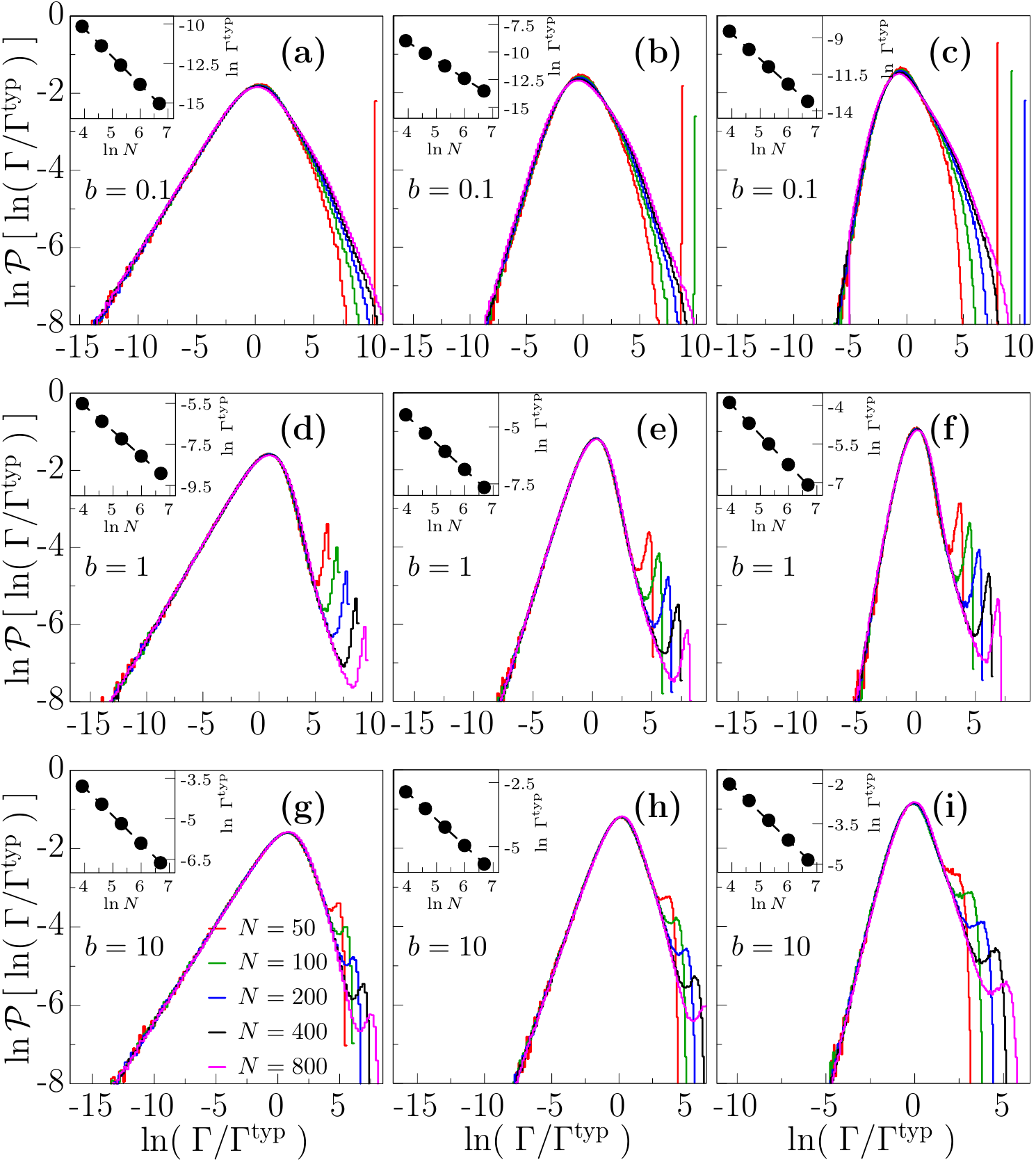}
\caption{Distribution of logarithm of resonance widths, normalized to its typical value, for the PBRM model at criticality with $\beta=1, 2,$ and 4, in the first, second, and third columns, respectively. The system sizes $N$ are indicated in panel (g). The values of the bandwidth are indicated in each panel. The insets show the scaling law $\Gamma^\text{typ}\propto N^{-\lambda}$. The black dashed lines are the best fitting of the scaling law to the numerical data. The error bars are smaller than the symbols.}
\label{fig:resonances}
\end{figure}

The logarithm of the distribution of the resonance widths, normalized to its typical value $\Gamma^\text{typ}\equiv\exp\bra\ln\Gamma\ket$, for the PBRM model at criticality with $\beta=1$ (first column), 2 (second column), and 4 (third column) are shown in Fig.~\ref{fig:resonances}. The wire lengths under consideration are indicated in Fig.~\ref{fig:resonances}(g), and the different values of $b$ in each panel. For the histograms, only 25\% of the eigenvalues around the center of the spectrum, $E=0$, are used. The typical value $\Gamma^\text{typ}$ follows a power-law with respect to the wire length $N$, $\Gamma^\text{typ}\propto N^{-\lambda}$, as observed in the insets of the same figure. There, the dashed lines are the best fittings to the numerical data. The resulting exponents $\lambda$ for each case are reported in Table~\ref{tab:scalingsVsN}.


\subsection{Wigner delay time vs spectral and eigenstate properties of the isolated PBRM model}

It is well known that the spatial fluctuations of the eigenstates of disordered systems at criticality show multifractal behavior~\cite{Janssen1994,Janssen1998,Evers2008,Hashimoto2008,Faez2009,Richardella2010}. This behavior is characterized by a set of \emph{generalized dimensions} $D_{q}$ or \emph{multifractal dimensions}, where $q$ is a real number. Furthermore, the multifractal properties of the eigenstates can also be studied through the Wigner delay time, as shown below.

For disordered systems at criticality, a relationship between the inverse moments of the Wigner delay time and the multifractal dimensions of the eigenstates of the corresponding isolated system $D_{q}$ is given by~\cite{Ossipov2005,Mendez2005,Mirlin2006},
\begin{equation}
\label{eq:WdtInvDq}
\bra\wdt^{-q}\ket\propto N^{-\sigma_{q}},\quad\text{where}\quad
\sigma_{q}\equiv q\;D_{q+1}
\end{equation}
and for the PBRM model at criticality the following functional form for $\sigma_{q}$ as a function of the bandwidth $b$ has been proposed~\cite{Mendez2014}
\begin{equation}
\label{eq:sigmaqofb}
\sigma_q\approx\dfrac{q}{1+(\alpha_{q+1}b)^{-1}}
\end{equation}
with $\alpha_q$ being fitting constants.
In addition, for PBRM models at criticality, the typical value of the Wigner delay time, defined as $\wdt^\text{typ}\equiv\exp\bra\ln\wdt\ket$, obeys the scaling law given by~\cite{Mendez2006}
\begin{equation}
\label{eq:WdttypDq}
\wdt^\text{typ}\propto N^{\sigma_{\tau}},\quad\text{where}\quad
\sigma_{\tau}\equiv D_{1}.
\end{equation}
%

%
%

%
\begin{figure}
\centering
\includegraphics[width=0.48\textwidth]{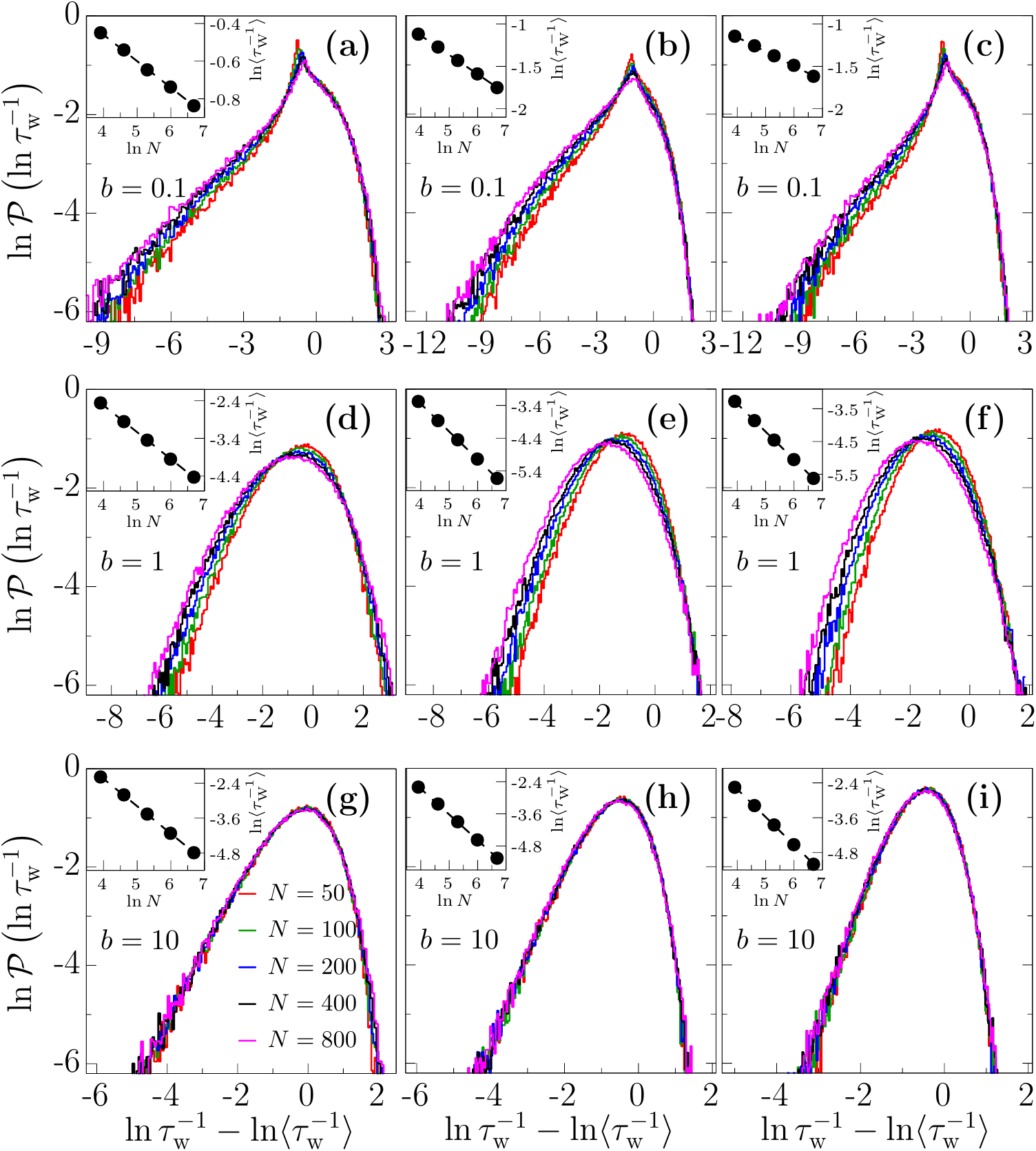}
\caption{Distribution of the first inverse moment of the Wigner delay time for the PBRM model at criticality in the presence of the $\beta=1$ (first column), 2 (second column), and 4 (third column) symmetries. In top panels $b=0.1$, in middle panels $b=1$, and in bottom panels $b=10$. The wire lengths $N$ are indicated in panel (g). The insets show the best fitting (black dashed lines) using the scaling law~(\ref{eq:WdtInvDq}) to the numerical data (symbols). The error bars are smaller than the symbols size.}
\label{fig:lnPlnWdtinv}
\end{figure}
%

%
\begin{figure}
\centering
\includegraphics[width=0.48\textwidth]{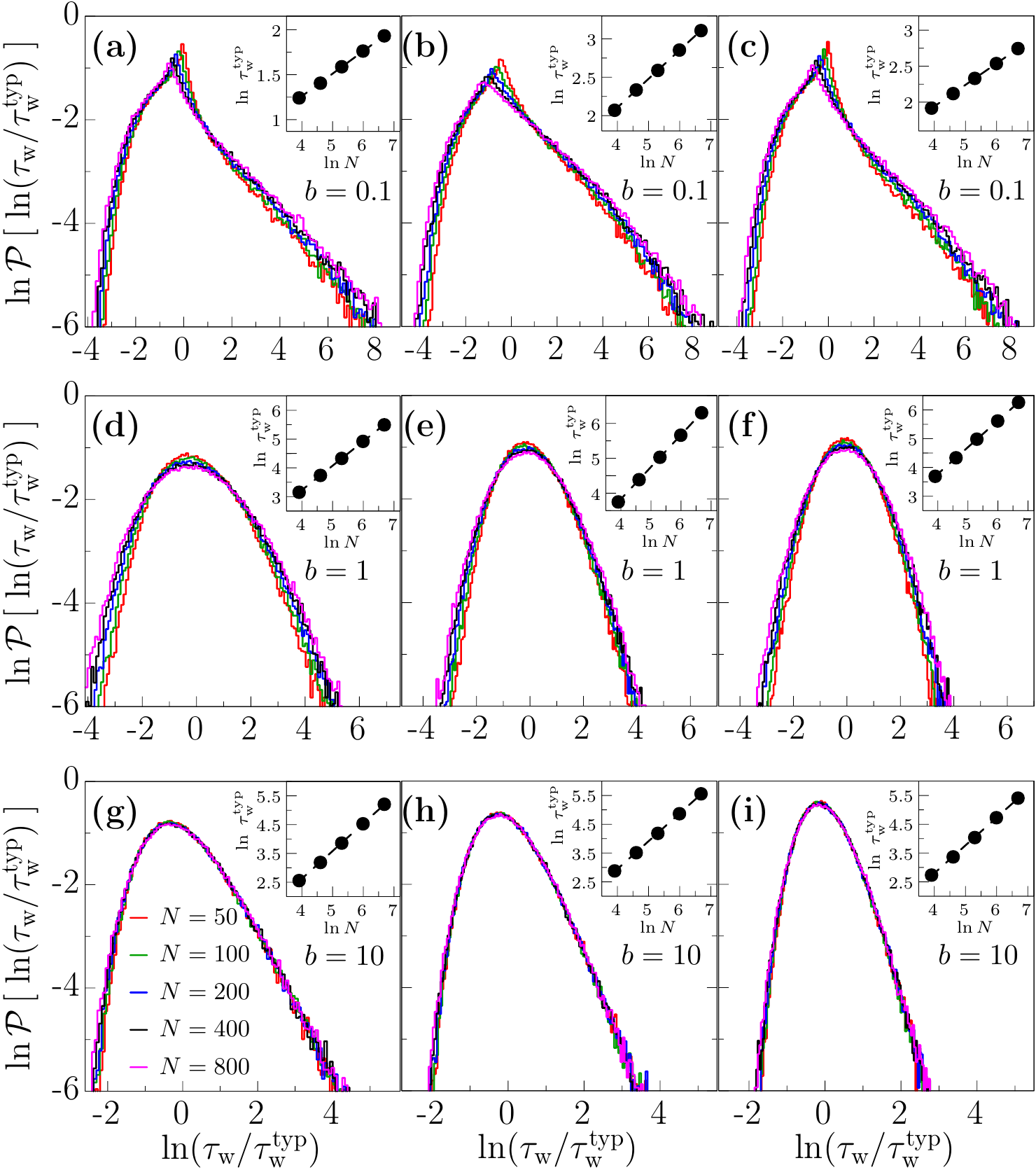}
\caption{Distribution of the Wigner delay time normalized to its typical value for the PBRM model at criticality. The symmetry classes are $\beta=1$ (first column), 2 (second column), and 4 (third column). The bandwidth $b$ is indicated in the panels and the wire lengths $N$ are indicated in panel (g). The insets show fittings of the scaling law~(\ref{eq:WdttypDq}) (dashed lines) to the numerical data (dots). The error bars are smaller than the symbol size.}
\label{fig:lnPlnWdttyp}
\end{figure}

For completeness, the \emph{level compressibility} $\chi$, a quantity often used to characterize the fluctuations of spectra of disordered systems at criticality, is also analyzed. In the metallic [insulator] regime $\chi=0$ [$\chi=1$] while at intermediate regimes (neither metallic nor insulator) $0<\chi<1$~\cite{Chalker1996,Klesse1997}. For PBRM models in the presence of $\beta=1$ and 2 symmetries, the level compressibility as a function of $b$ is given by~\cite{Evers2008,Kravtsov2011}
\begin{equation}
\label{eq:chi}
\chi=
\left\{
\begin{array}{lr}
1 - 4b,~~~~~~~~~~~~\;~~~~~~~~~~~~~~~~\beta=1, & b\ll1 , \\
1-\pi\sqrt{2}b+\frac{4}{3}(2-\sqrt{3})\pi^2 b^2, \, \, \, \beta=2 , & b\ll1 , \\
1/2\beta\pi b, &  b\gg1 .
\end{array} \right.
\end{equation}
In addition a heuristic relation between $\chi$ and $\sigma_{q}$ is also known, namely~\cite{Mendez2014}
\begin{equation}
\label{eq:chiofsigmaq}
\chi\approx\frac{q-\sigma_{q}}{q(\sigma_{q}+1)}.
\end{equation}

In Fig.~\ref{fig:lnPlnWdtinv}, the distribution of the logarithm of the first inverse moment of the Wigner delay time for the PBRM model at criticality is reported. The values of the bandwidth $b$ and wire lengths $N$ are indicated in the panels. The insets show $\ln\langle\wdt^{-1}\rangle$ vs $\ln N$ as dots while the dashed lines correspond to fittings to the numerical data with Eq.~(\ref{eq:WdtInvDq}). The resulting exponents $\sigma_{q}$ from the fittings are reported in Table~\ref{tab:scalingsVsN}.  A good agreement between the numerical data and the scaling law of Eq.~(\ref{eq:WdtInvDq}) is observed.

The behavior of the logarithm of the distribution of $\ln \big( \tau_\text{W}/\tau_\text{W}^\text{typ} \big)$ for the PBRM model at criticality with the three symmetry classes $\beta=1, 2,$ and 4, for different wire lengths $N$ [see 
Fig.~\ref{fig:lnPlnWdtinv}(g)], and different bandwidths $b$, is plotted in 
Fig.~\ref{fig:lnPlnWdttyp}. The insets show the logarithm of $\wdt^\text{typ}$ as a function of $\ln N$. The dots correspond to numerical results while the dashed lines correspond to fittings with the scaling law of Eq.~(\ref{eq:WdttypDq}) to the numerical data. The exponents $\sigma_{\tau}$ resulting from the fittings are reported in Table~\ref{tab:scalingsVsN}. A good agreement between the numerical data and the scaling law~(\ref{eq:WdttypDq}) is observed.

%
\begin{table*}
\caption{ Values of the parameters obtained from the power-law fittings to the data of the insets in Figs.~\ref{fig:resonances}, \ref{fig:lnPlnWdtinv}, and \ref{fig:lnPlnWdttyp}.}
%
%
%
%
%

\begin{ruledtabular}
\begin{tabular}{cccccc} 
\multicolumn{3}{c}{}      & %
 FIG.~\ref{fig:resonances}  &
 FIG.~\ref{fig:lnPlnWdtinv} & %
 FIG.~\ref{fig:lnPlnWdttyp} \\ \hline%
                                                                      &     &             & {\boldmath{$\Gamma^\text{typ}\propto N^{-\lambda}$}} 
                                                                                          & {\boldmath{$\langle\tau_\text{W}^{-1}\rangle\propto N^{-\sigma_{q}}$}}  
                                                                                          & {\boldmath{$\tau^\text{typ}\propto N^{\sigma_{\tau}}$}}                    \\ 
                                                                      & $b$ & Inset panel & $\lambda$             &  $\sigma_{q}$         & $\sigma_{\tau}$            \\ \hline
\parbox[t]{2mm}{\multirow{3}{*}{\rotatebox[origin=c]{90}{$\beta=1$}}} & 0.1 & (a)         & $1.7647 \pm 0.0072$   &  $0.1404 \pm 0.0037$  & $0.2507 \pm 0.0048$        \\ 
                                                                      & 1~~ & (d)         & $1.2265 \pm 0.0080$   &  $0.7133 \pm 0.0068$  & $0.8465 \pm 0.0055$        \\
                                                                      & 10  & (g)         & $1.0276 \pm 0.0213$   &  $0.9439 \pm 0.0180$  & $0.9586 \pm 0.0163$        \\ \hline
\parbox[t]{2mm}{\multirow{3}{*}{\rotatebox[origin=c]{90}{$\beta=2$}}} & 0.1 & (b)         & $1.6424 \pm 0.0036$   &  $0.2282 \pm 0.0034$  & $0.3744 \pm 0.0035$        \\ 
                                                                      & 1~~ & (e)         & $1.1490 \pm 0.0063$   &  $0.8469 \pm 0.0023$  & $0.9217 \pm 0.0027$        \\
                                                                      & 10  & (h)         & $1.0187 \pm 0.0244$   &  $0.9609 \pm 0.0192$  & $0.9685 \pm 0.0204$        \\ \hline
\parbox[t]{2mm}{\multirow{3}{*}{\rotatebox[origin=c]{90}{$\beta=4$}}} & 0.1 & (c)         & $1.7223 \pm 0.0245$   &  $0.1699 \pm 0.0052$  & $0.2990 \pm 0.0023$        \\ 
                                                                      & 1~~ & (f)         & $1.1596 \pm 0.0068$   &  $0.8499 \pm 0.0060$  & $0.9257 \pm 0.0040$        \\
                                                                      & 10  & (i)         & $1.0218 \pm 0.0456$   &  $0.9590 \pm 0.0175$  & $0.9680 \pm 0.0196$        \\ 
\end{tabular}
\end{ruledtabular}


\label{tab:scalingsVsN}
\end{table*}

\begin{figure}
\centering
\includegraphics[width=0.48\textwidth]{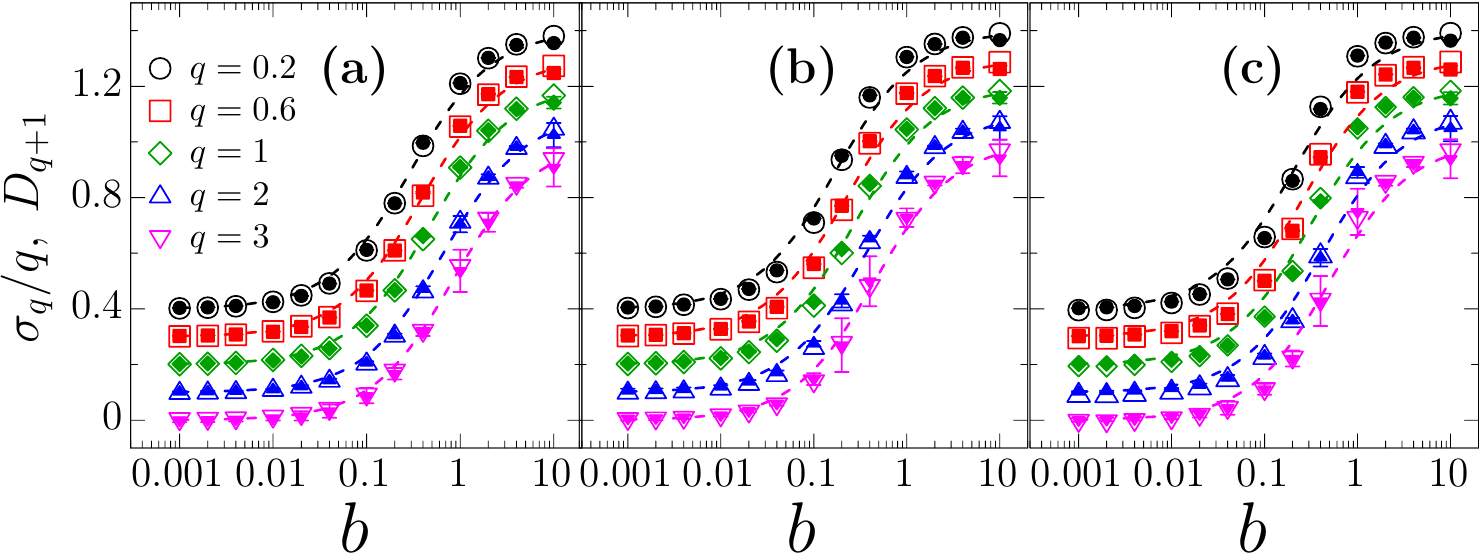}
\caption{Multifractal dimensions $D_{q}$ for the PBRM model at criticality as a function of the bandwidth $b$. The empty symbols are obtained from direct diagonalization of the closed system~(\ref{eq:Hij}). The filled symbols correspond to results obtained from the scaling law~(\ref{eq:WdtInvDq}). The dashed lines correspond to the heuristic relation~(\ref{eq:sigmaqofb}). The error bars are the rms of the residuals. For the sake of clarity, the symbols (lines) are displaced vertically upward.}
\label{fig:WdtandDq}
\end{figure}

In Fig.~\ref{fig:WdtandDq}, the multifractal dimension $D_{q}$ as a function of the bandwidth $b$ for several values of $q$ is plotted. Figures~\ref{fig:WdtandDq}(a), \ref{fig:WdtandDq}(b), and \ref{fig:WdtandDq}(c) correspond to the system in the presence of the $\beta=1, 2,$ and 4 symmetry, respectively. The empty symbols are computed by direct diagonalization of the isolated PBRM mode at criticality (see also Ref.~\cite{Carrera2021} for more details). The filled symbols are obtained from the scaling law~(\ref{eq:WdtInvDq}) and the dashed lines correspond to Eq.~(\ref{eq:sigmaqofb}) with the $\alpha_{q+1}$ taken from Ref.~\cite{Carrera2021}. These values of $\alpha_{q+1}$ are: 3.33 (5.70, 4.82), 2.55 (4.45, 3.82), 2.11 (3.73, 3.22), 1.52 (2.74, 2.41), and 1.21 (2.22, 1.98) for $\beta=1\;(2,4)$ and $q=0.2, 0.6, 1, 2$, and 3, respectively~\cite{Carrera2021}. 
For the three symmetry classes $\beta=1$ (a), 2 (b), and 4 (c) a good correspondence between the direct calculation of $D_{q}$ (empty symbols) and the analytics (dashed lines) is found. These results show that for the one channel setup, the multifractal properties of the isolated PBRM model at criticality can be directly extracted from Wigner delay time, a transport property. This is convenient from an experimental point of view.

%

\begin{figure}
\centering
\includegraphics[width=0.48\textwidth]{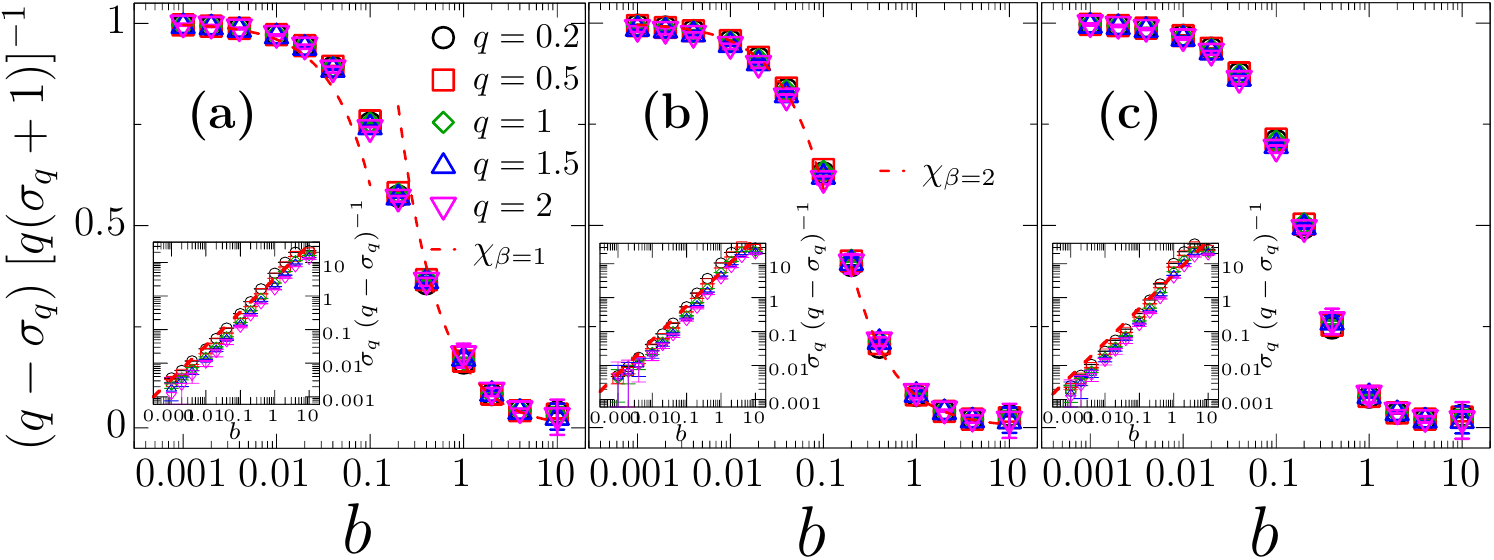}
\caption{Level compressibility of the PBRM model at criticality as a function of the bandwidth $b$. The symmetries are $\beta=1, 2,$ and 4, in panels (a), (b), and (c), respectively. The symbols are obtained from Eq.~(\ref{eq:chiofsigmaq}), the dashed red lines correspond to expression~(\ref{eq:chi}). Inset: Recursive relation~(\ref{eq:recursivesigmaq}), the dashed red lines ($\sim b$) are plotted to guide the eye. The error bars are the rms of the residuals. }
\label{fig:chiofb}
\end{figure}

In Fig.~\ref{fig:chiofb} the spectral compressibility of the PBRM model at criticality is reported. The symbols are obtained from Eq.~(\ref{eq:chiofsigmaq}) with the $\sigma_{q}$ extracted from the scaling law~(\ref{eq:WdtInvDq}). The dashed lines correspond to the theoretical prediction of Eq.~(\ref{eq:chi}). For the PBRM model in the presence of the symmetry classes, $\beta=1$ and 2, 
Figs.~\ref{fig:chiofb}(a) and ~\ref{fig:chiofb}(b), respectively, a good agreement with the analytics is observed. For the $\beta=4$ case, there is no theoretical prediction available to compare with. However, the following recursive relation~\cite{Mendez2014}:
\begin{equation}
\label{eq:recursivesigmaq}
\sigma_{q}(q-\sigma_{q})\approx\alpha_{q+1}b
\end{equation}
is known. Then, the PBRM model in the $\beta=4$ case can also be contrasted with this last relation. For this purpose, in the insets of Fig.~\ref{fig:chiofb}, the recursive relation~(\ref{eq:recursivesigmaq}) (symbols) is shown.
The dashed lines are $\sim b$. A good agreement with the numerical data (symbols) is observed.

%

To this point, the analysis of the PBRM model at criticality with one open 
channel has been performed. In the next sections, the setup in which the scattering system supports $M$ open channels is studied and contrasted with available RMT predictions.


\section{PBRM model with \boldmath $M$ open channels}
\label{sec:MChannels}

In this section, the analysis of the scattering and transport properties of the PBRM model at criticality with $M$ open channels is presented. As in the previous section, this is performed following the scattering matrix approach of 
Eq.~(\ref{eq:scatteringmatrix2}). As shown in Sec.~\ref{sec:onesinglechannel} for the one open-channel setup, the scattering and transport properties of the PBRM model at criticality do not change significantly with the wire length $N$. A similar behavior has also been found for the $M$ open-channels setup in the $\beta=1$ case~\cite{Mendez2005,Mendez2006,Mendez2010}. Here, this has also been verified for the $\beta=2$ and 4 cases (not shown). Therefore, in what follows the wire length is set to $N=50$. For the statistical analysis, $10^{6}$ realizations of the scattering matrix $S$ in the perfect coupling regime are considered. In the same line as in Sec.~\ref{sec:onesinglechannel}, known analytical RMT predictions are presented first and later they are compared with numerical simulations in the appropriate limits.


\subsection{Scattering properties}

Within the RMT approach of quantum transport, the average of the magnitude of the elements of the scattering matrix $S$ has been obtained~\cite{Beenakker1997}. That is
\begin{equation}
\label{eq:SnmRMT}
\bra |S_{nm}|^{2} \ket_\rmt = 
 \frac{1 - (1 - 2/\beta) \delta_{n,m}}{M -1 + 2/\beta} ,
\end{equation}
where $\beta$ is the symmetry class present in the system, $M$ is the number of open channels, and $\delta_{nm}$ is the usual Kronecker delta.

From Eq.~(\ref{eq:SnmRMT}) and based on numerical simulations for the $M = 2$ open-channel case, it has been conjectured that the average of the $S$-matrix elements, $\langle |S_{12}|^{2} \rangle$ and $\langle |S_{11}|^{2} \rangle$, can be expressed as a function of the bandwidth $b$~\cite{Alcazar2009,Mendez2010}. These are given by
\begin{align}
\label{eq:S12ofb}
\bra\vert S_{12}\vert^{2}\ket(b) &= 
\dfrac{\langle |S_{12}|^{2} \rangle_\rmt}{1+(\varepsilon b)^{-2}} \quad \mathrm{and} \\
\label{eq:S11ofb}
\bra\vert S_{11}\vert^{2}\ket(b) &= 1 - \bra\vert S_{12}\vert^{2}\ket ,
\end{align}
where $\varepsilon$ is a free parameter to be determined by the best fitting to the numerical data. Furthermore, by using a phenomenological expression that relates the bandwith $b$ with the correlation dimension $D_{2}$, which is a broadly accepted measure of the spatial extension of the eigenfunctions of disordered systems at the critical point, Eqs.~(\ref{eq:S12ofb}) and~(\ref{eq:S11ofb}) can also be written in terms of $D_{2}$~\cite{Alcazar2009,Mendez2010}. For the $\beta=1$ case one gets
\begin{align}
\label{eq:S12ofD2beta1}
\bra\vert S_{12}\vert^{2}\ket(D_{2}) &= 
\dfrac{\langle |S_{12}|^{2} \rangle_\rmt}{1+(\kappa/\varepsilon)^{2}\;(D_{2}^{-1}-1)^{2}} \quad \mathrm{and} \\
\label{eq:S11ofD2beta1}
\bra\vert S_{11}\vert^{2}\ket(D_{2}) &= 
1-\bra\vert S_{12}\vert^{2}\ket(D_{2}),
\end{align}
where $\kappa$ is a fitting parameter and $\varepsilon$ is obtained from Eq.~(\ref{eq:S12ofb}). Meanwhile, for the $\beta=2$ and 4 cases one gets
\begin{align}
\label{eq:S12ofD2beta24}
\bra\vert S_{12}\vert^{2}\ket(D_{2})&=
\left\{
\begin{array}{lr}
\dfrac{\bra\vert S_{12}\vert^{2}\ket_\rmt}{1 + \left(\kappa D_{2}/\pi\right)^{-2}}, & b\ll1 , \\[0.5cm]
\dfrac{\bra\vert S_{12}\vert^{2}\ket_\rmt}{1 + \left[ 2\pi( 1-D_{2})/\rho \right]^{2}}, & b\gg1 ,
\end{array} \right.\\[0.5cm]
\label{eq:S11ofD2beta24}
\bra\vert S_{11}\vert^{2}\ket(D_{2}) &= 
1-\bra\vert S_{12}\vert^{2}\ket(D_{2}),
\end{align}
where $\kappa$ and $\rho$ are fitting parameters, different for each symmetry class $\beta$. Notice that the expressions~(\ref{eq:S12ofb}), (\ref{eq:S12ofD2beta1}), and (\ref{eq:S12ofD2beta24}) were obtained heuristically and its analytical derivation remains to be proven, while the relations~(\ref{eq:S11ofb}), (\ref{eq:S11ofD2beta1}), and (\ref{eq:S11ofD2beta24}) are a consequence of the flux conservation condition of the $S$ matrix. However, as shown below, these equations describe well the numerical data.

\begin{figure}
\centering
\includegraphics[width=0.48\textwidth]{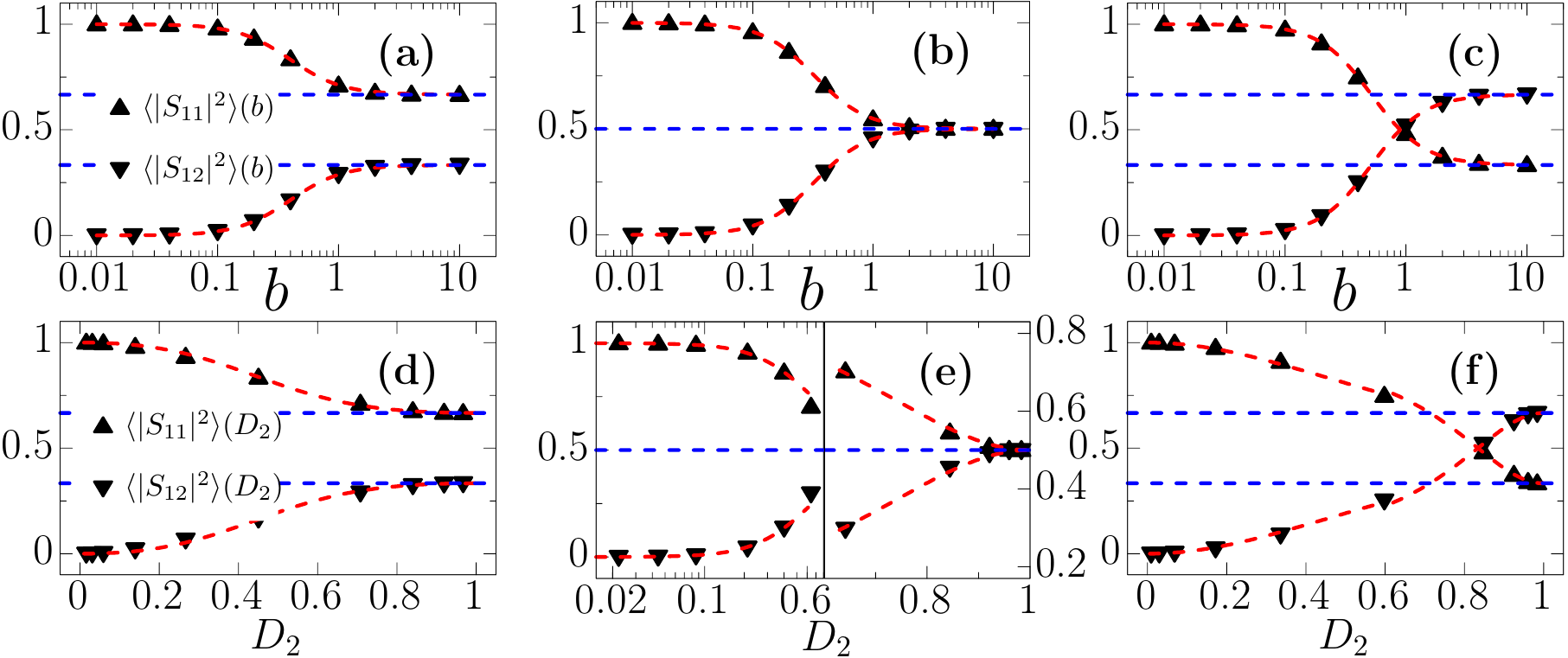}
\caption{Average of the $S$-matrix elements for the PBRM model at criticality with $M=2$ open channels. The symmetries are $\beta=1$ (first column), 2 (second column), and 4 (third column). The symbols correspond to numerical simulations, the dashed blue lines are the corresponding RMT prediction~(\ref{eq:SnmRMT}). Top panels: Average of the $S$-matrix elements as a function of the bandwidth $b$. The dashed red lines are Eq.~(\ref{eq:S12ofb}) and~(\ref{eq:S11ofb}). Bottom panels: same as top panels but as a function of the correlation dimension $D_{2}$ compared with 
Eqs.~(\ref{eq:S12ofD2beta1})-(\ref{eq:S11ofD2beta24}). Error bars are smaller than the symbol size. }
\label{fig:S12S11}
\end{figure}

In Fig.~\ref{fig:S12S11} the average of the $S$-matrix elements, when the PBRM model at criticality supports $M=2$ open channels, is shown for the $\beta=1, 2,$ and 4 symmetries, in the first, second, and third columns, respectively. The symbols correspond to numerical simulations and the dashed blue lines are the RMT prediction~(\ref{eq:SnmRMT}) for which $\bra |S_{12}|^{2} \ket_\rmt =1/3,1/2,$ and $2/3$ for $\beta=1, 2,$ and 4, respectively. In the same figure, $\bra |S_{11}|^{2} \ket_\rmt=1-\bra |S_{12}|^{2} \ket_\rmt$ is also shown. The dashed red curves in Figs.~\ref{fig:S12S11}(a), \ref{fig:S12S11}(b), and 
\ref{fig:S12S11}(c) correspond to the generalizations of Eqs.~(\ref{eq:S12ofb}) and (\ref{eq:S11ofb}) for the $\beta=1, 2,$ and 4 symmetry, respectively. 
Figures~\ref{fig:S12S11}(d), \ref{fig:S12S11}(e), and \ref{fig:S12S11}(f), show $\bra\vert S_{12}\vert^{2}\ket(D_{2})$ with the values of $D_{2}$ extracted from Fig.~\ref{fig:WdtandDq}. The dashed red lines are Eqs.~(\ref{eq:S12ofD2beta1})-(\ref{eq:S11ofD2beta24}). The obtained fitting parameters $\varepsilon$, $\kappa$, and $\rho$, for each symmetry class are reported in Table~\ref{tab:fittingpar}. In all the cases, it is observed that for large $b$ and $D_{2}$ the model is well described by the RMT predictions while the corresponding generalizations work well for any $b$ and $D_{2}$.

%
%

%
\begin{table}
\caption{\label{tab:fittingpar} Values of the parameters obtained by fittings to the data of figure~\ref{fig:S12S11}.}
%
%
%
%
%

\begin{ruledtabular}
\begin{tabular}{cccc}
$\beta$   & $\varepsilon$       & $\kappa$            & $\rho$              \\ \hline%
1        & $2.5129 \pm 0.0017$ & $2.1043 \pm 0.0242$ &                      \\ 
2        & $3.0618 \pm 0.0022$ & $4.7867 \pm 0.0017$ & $2.7857 \pm 0.0065$  \\
4        & $1.9359 \pm 0.0030$ & $3.9909 \pm 0.0078$ & $1.8036 \pm 0.0038$  
\end{tabular}
\end{ruledtabular}


\end{table}


\subsection{Transmission and shot noise power}

In this section, known analytical results from RMT for the scattering and transport properties of complex media are first revised. Those results are expected to describe the PBRM model at criticality in the case of $b\gg1$, i.e., in the metallic-like regime. 

It is well established that given a scattering problem, the transmission coefficient can be obtained from the elements of the scattering matrix as
\begin{equation}
\label{eq:Defofconductance}
T=\text{Tr}(t\;t^{\dagger}),
\end{equation}
where Tr is the trace operation, $t$ is the transmission amplitude [see 
Eq.~(\ref{eq:scatteringmatrix2})], and the symbol $\dagger$ represents the adjoint of $t$.

For the $M=2$ open-channel setup, or two single-channel leads attached to a complex scattering media, the transmission distribution is given by~\cite{Beenakker1997}
\begin{equation}
\label{eq:PofTM=1}
\p(T) = \frac{1}{2} \beta T^{-1 + \beta/2}, \quad 0<T<1 ,
\end{equation}
for the symmetry class labeled by $\beta$.

Also, for the $M=4$ open-channel setup, or four single-channel leads attached to a complex scattering media, the distribution of $T$ for the $\beta=1$ case is~\cite{MelloBook}
\begin{eqnarray}
\label{eq:PofTB1M=2}
\p(T) & = & 
\left\{
\begin{array}{rl}
\frac{3}{2} T,  \qquad 0<T<1, \\ 
\frac{3}{2} (T - 2 \sqrt{T - 1}), \qquad 1<T<2 ,
\end{array}
\right. 
\end{eqnarray}
while for the $\beta=2$ case it is~\cite{MelloBook}
\begin{eqnarray}
\label{eq:PofTB2M=2}
\p(T) & = & 
\left\{
\begin{array}{rl}
2 T^{3},  \qquad 0<T<1, \\ 
2 (2 - T)^{3}, \qquad 1<T<2 .
\end{array}
\right.   
\end{eqnarray}
Now, for the $\beta=4$ symmetry, following Ref.~\cite{MelloBook} it is straightforward to arrive at
\begin{equation}
\label{eq:PofTB4M=2}
\p(T) =  
\left\{
\begin{array}{rl}
\frac{12}{7} T^{7},  \quad 0<T<1, \\ [0.2cm]
\frac{12}{7} (2-T)^{5} (T^{2} + 10 T -10), \quad 1<T<2 .
\end{array}
\right .
\end{equation}
In general, the tail of the transmission distribution for all $\beta$ and any number of open channels $M$ decays as~\cite{MelloBook}
\begin{equation}
\label{eq:tailPofT}
\p(T) \propto T^{-1 + \beta M^{2}/2}.
\end{equation}

Furthermore, for a complex scattering media attached to two leads supporting respectively $N_{1}$ and $N_{2}$ open channels, the average transmission and its variance are given by~\cite{Baranger1994,Beenakker1997}
\begin{equation}
\label{eq:avgT}
\left \langle T \right \rangle = 
                  \frac{N_{1} N_{2}}{N_{1} + N_{2} -1 + 2/\beta}
\end{equation}
and
\begin{widetext}
\begin{equation}
\label{eq:avgVarT}
\mathrm{var} \left( T \right) = \frac{2 N_{1} N_{2} (N_{1} - 1 + \frac{2}{\beta}) (N_{2} - 1 + \frac{2}{\beta})}{\beta (N_{1} + N_{2} - 2 + \frac{2}{\beta}) (N_{1} + N_{2} - 1 + \frac{4}{\beta}) (N_{1} + N_{2} - 1 + \frac{2}{\beta})^{2}} ,
\end{equation}
\end{widetext}
for all symmetry classes labeled by $\beta$.

Another transport quantity of interest is the so-called shot noise power $P$, defined as $P=\bra\text{tr}\left( tt^{\dagger}-tt^{\dagger}\;tt^{\dagger}  \right) \ket$, whose probability distribution is given by~\cite{Savin2006}
\begin{equation}
\label{eq:shotnoise}
P=\frac{
N_{1}\left(N_{1}-1+\frac{2}{\beta} \right) N_{2} 
     \left(N_{2}-1+\frac{2}{\beta} \right)
        }{
     \left(K - 2 + \frac{2}{\beta} \right)
     \left(K - 1 + \frac{2}{\beta} \right)
     \left(K - 1 + \frac{4}{\beta} \right)
        },
\end{equation}
where $K=N_{1}+N_{2}$ is the total number of open channels. Note that although the distribution~(\ref{eq:shotnoise}) was derived for the $\beta=1$ and 2 symmetries, it also encompasses the symplectic case, $\beta=4$, as will be verified below for the PBRM model with $b\rightarrow\infty$.

%
\begin{figure}
\centering
\includegraphics[width=0.48\textwidth]{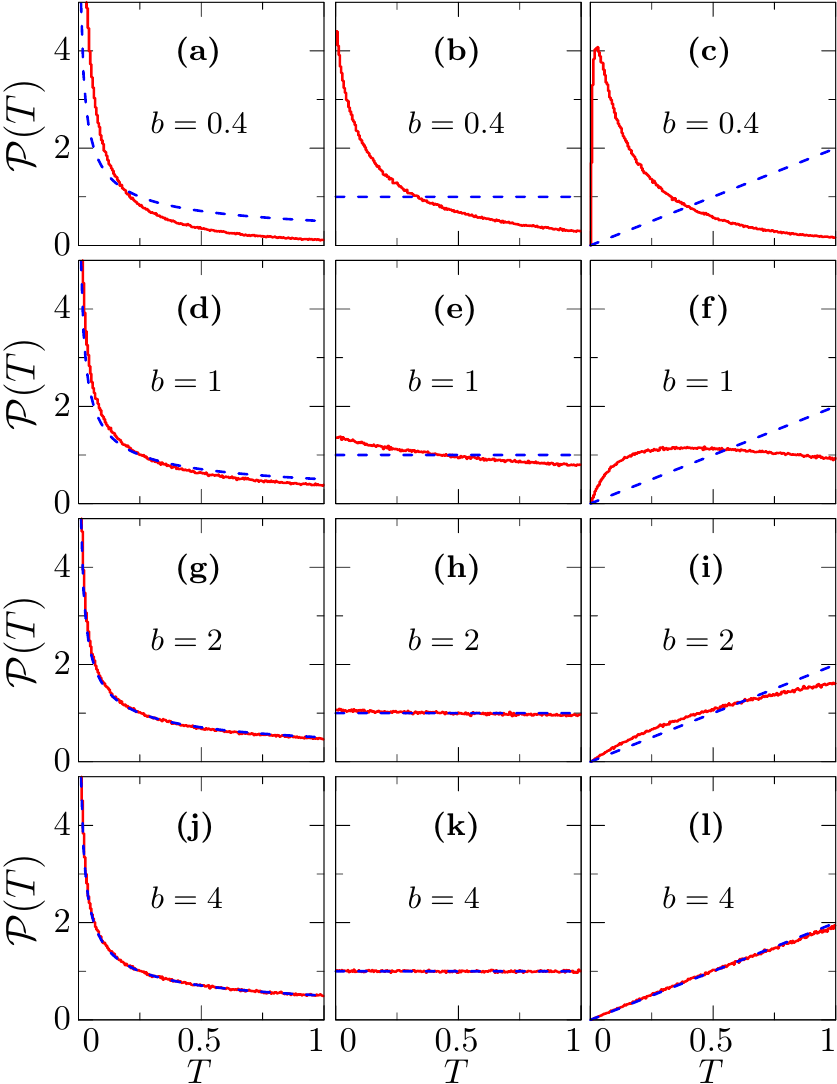}
\caption{Transmission distribution of the PBRM model at criticality with $M=2$ open channels and several values of $b$. The symmetries are $\beta=1, 2,$ and 4, in the first, second, and third columns, respectively. The red lines correspond to histograms obtained from numerical simulations while the dashed blue lines are the RMT prediction~(\ref{eq:PofTM=1}) for the respective $\beta$. A smooth transition from localized-like to metallic-like regime is observed for the three symmetry classes.}
\label{fig:PofTM=1}
\end{figure}

In Eqs.~(\ref{eq:avgT})-(\ref{eq:shotnoise}) the number of open channels, $N_{1}$ and $N_{2}$, may be different. However, in order to compare these quantities with numerical simulations of the PBRM model, we set $N_{1} = N_{2}$ with $M = N_{1} + N_{2}$ the total number of open channels.

For the PBRM model at criticality with two single-channel leads attached to it ($M=2$ open channels), the transmission distribution for several values of $b$ and $\beta=1, 2,$ and 4 (first, second, and third column, respectively), is shown in Fig.~\ref{fig:PofTM=1}. The red lines correspond to histograms obtained from numerical simulations while the dashed blue lines are Eq.~(\ref{eq:PofTM=1}) for the respective $\beta$. For the three symmetry classes under consideration, a smooth transition from a localized-like to a metallic-like regime is observed as the bandwidth $b$ increases. The metallic-like regime is reached when $b=4$ for which a good agreement with the RMT prediction~(\ref{eq:PofTM=1}) is obtained [see Figs.~\ref{fig:PofTM=1}(j), \ref{fig:PofTM=1}(k), and \ref{fig:PofTM=1}(l)].

%
\begin{figure}
\centering
\includegraphics[width=0.48\textwidth]{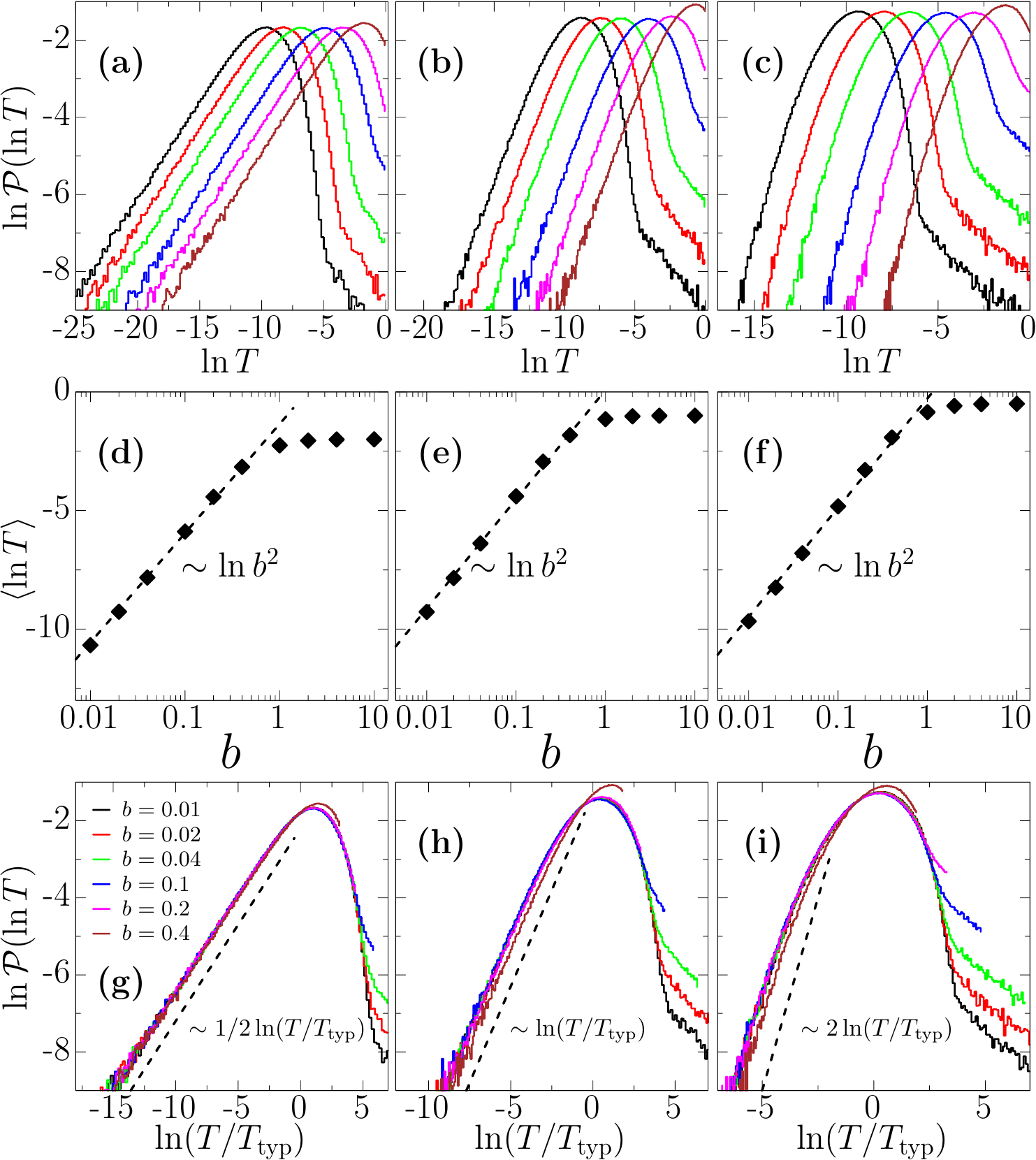}
\caption{Distribution of the logarithm of transmission for the PBRM model at criticality with $M=2$ and several values of the bandwidth $b$, as indicated in panel (g). The symmetries are $\beta=1$ (first column), 2 (second column), and 4 (third column). The error bars in panels (d)-(f) are smaller than the symbol size. See text for the discussion. }
\label{fig:lnP_lnT_M=1}
\end{figure}

It is instructive to look at the behavior of the transmission distribution $\mathcal{P}(T)$ in the insulator-like regime $b\ll1$ where RMT predictions are not available. In this regime where $\mathcal{P}(T)\approx0$ it is more convenient to analyze the distribution of $\ln T$. For the PBRM model at criticality with $M=2$, the distribution of the logarithm of $T$ for $\beta=1$ (first column), 2 (second column) and 4 (third column) and several values of the $b<1$, as indicated in Fig.~\ref{fig:lnP_lnT_M=1}(g), is shown in Fig.~\ref{fig:lnP_lnT_M=1}. 
In Figs.~\ref{fig:lnP_lnT_M=1}(a), \ref{fig:lnP_lnT_M=1}(b), and \ref{fig:lnP_lnT_M=1}(c), it is observed that the shape and width of $\ln\p(\ln T)$ do not change despite the fact that $b$ varies. This means that $\ln\p(\ln T)$ should be scale invariant, a property that was confirmed before for $\beta=1$ with the typical transmission $T^{\text{typ}}=\exp\bra\ln T \ket$ as scaling parameter. Also note in Figs.~\ref{fig:lnP_lnT_M=1}(d)--(f) that $\bra\ln T\ket\sim \ln b$ for $b<1$. Indeed, all distributions $\ln\p(\ln T)$ fall on top of each other when plotted as a function of $\ln (T/T^{\text{typ}})$, as shown in 
Figs.~\ref{fig:lnP_lnT_M=1}(g), \ref{fig:lnP_lnT_M=1}(h), and \ref{fig:lnP_lnT_M=1}(i). We stress that this behavior has previously been reported for the PBRM model in the presence of the $\beta=1$ symmetry but neither for the $\beta=2$ (periodic) nor for the $\beta=4$ symmetries.

Since the tails of the distribution $\p(T)$ from the PBRM model in the metallic-like regime ($b\rightarrow\infty$) are expected to correspond to Eq.~(\ref{eq:tailPofT}), then it is reasonable to assume that those tails may be described by
\begin{equation}
\label{eq:tailPofTPBRM}
\p(T)\propto T^{\nu}
\end{equation}
for all $b$, with $\nu$ a fitting parameter. Furthermore, for the PBRM model with $\beta=1$ symmetry, in Ref.~\cite{Antonio2009} the following relationship between the exponent $\nu$ and the correlation dimension $D_{2}$:
\begin{equation}
\label{eq:nuofD2}
\nu(D_{2})\propto(1-D_{2})^{2},\quad b\gg1,
\end{equation}
has been proposed.

%
\begin{figure}
\centering
\includegraphics[width=0.48\textwidth]{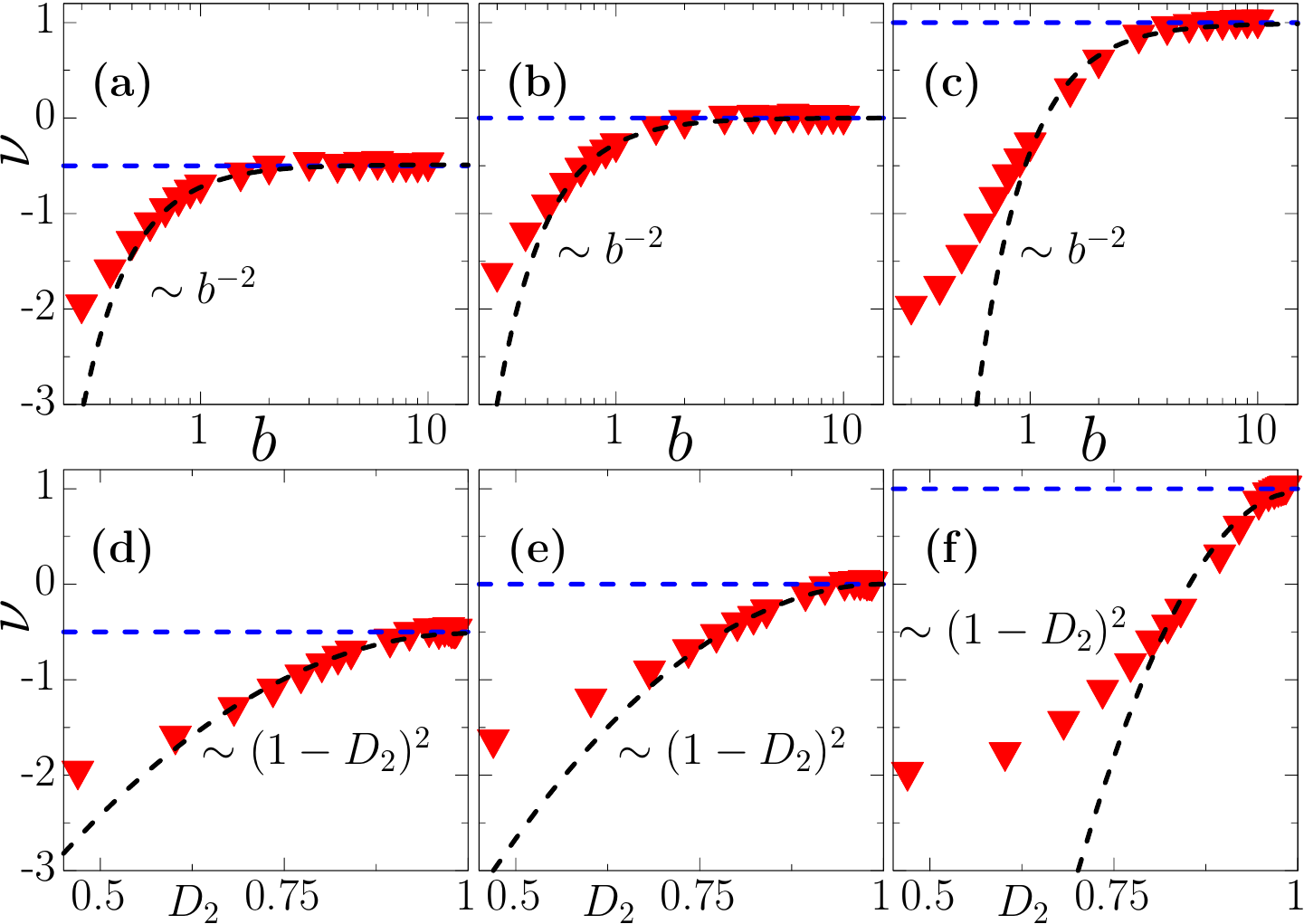}
\caption{Exponents $\nu$ extracted from the best fitting of Eq.~(\ref{eq:tailPofTPBRM}) to the numerical transmission distribution of the PBRM model at criticality for $M=2$. The symmetries are $\beta=1, 2,$ and 4, in the first, second, and third columns respectively. Top panels: $\nu$ as a function of the bandwidth $b$. The red inverted triangles correspond to numerical results for $T>0.5$, the dashed blue lines are the RMT prediction~(\ref{eq:tailPofT}) for the corresponding symmetry classes and the dashed black lines are the best fitting of $b^{2}$ to the data. Bottom panels: Same as top panels but as a function of the correlation dimension $D_{2}$ whose relationship with $b \gg 1$ is given through $D_{2}(b)=1-(2\pi b)^{-1}$~\cite{Evers2008,Antonio2009}. The fittings are performed by using $(1-D_{2})^{2}$. The error bars are the rms of the residual and are smaller than the symbol size. }
\label{fig:nuofbandD2}
\end{figure}

The asymptotic behavior of the transmission distribution for the PBRM model at criticality as a function of the bandwidth $b$ and of the correlation dimension $D_{2}$ is analyzed in Fig.~\ref{fig:nuofbandD2} for the $\beta=1, 2,$ and 4 symmetries in the first, second, and third columns, respectively. 
In Figs.~\ref{fig:nuofbandD2}(a), \ref{fig:nuofbandD2}(b), and \ref{fig:nuofbandD2}(c), for each value of $b$ the corresponding exponent $\nu$ (red inverted triangles) is obtained from the best fitting of Eq.~(\ref{eq:tailPofTPBRM}) to the numerical data for $T>0.5$. The exponent $\nu$ as a function of $D_{2}$ is shown in Figs.~\ref{fig:nuofbandD2}(d), \ref{fig:nuofbandD2}(e), and \ref{fig:nuofbandD2}(f). The relationship between $D_{2}$ and the bandwidth $b \gg 1$ is given by $D_{2}(b)=1-(2\pi b)^{-1}$~\cite{Evers2008,Antonio2009}. In all panels the dashed blue lines are the RMT prediction~(\ref{eq:tailPofT}) according to the symmetry class. The dashed black lines are the best fittings to the numerical data, which are proportional to $~b^{-2}$ (top panels) and to $(1-D_{2})^{2}$ (bottom panels). The latter shows that the relationship~(\ref{eq:nuofD2}) is valid for $D_{2}>0.75$.

%
\begin{figure}
\centering
\includegraphics[width=0.48\textwidth]{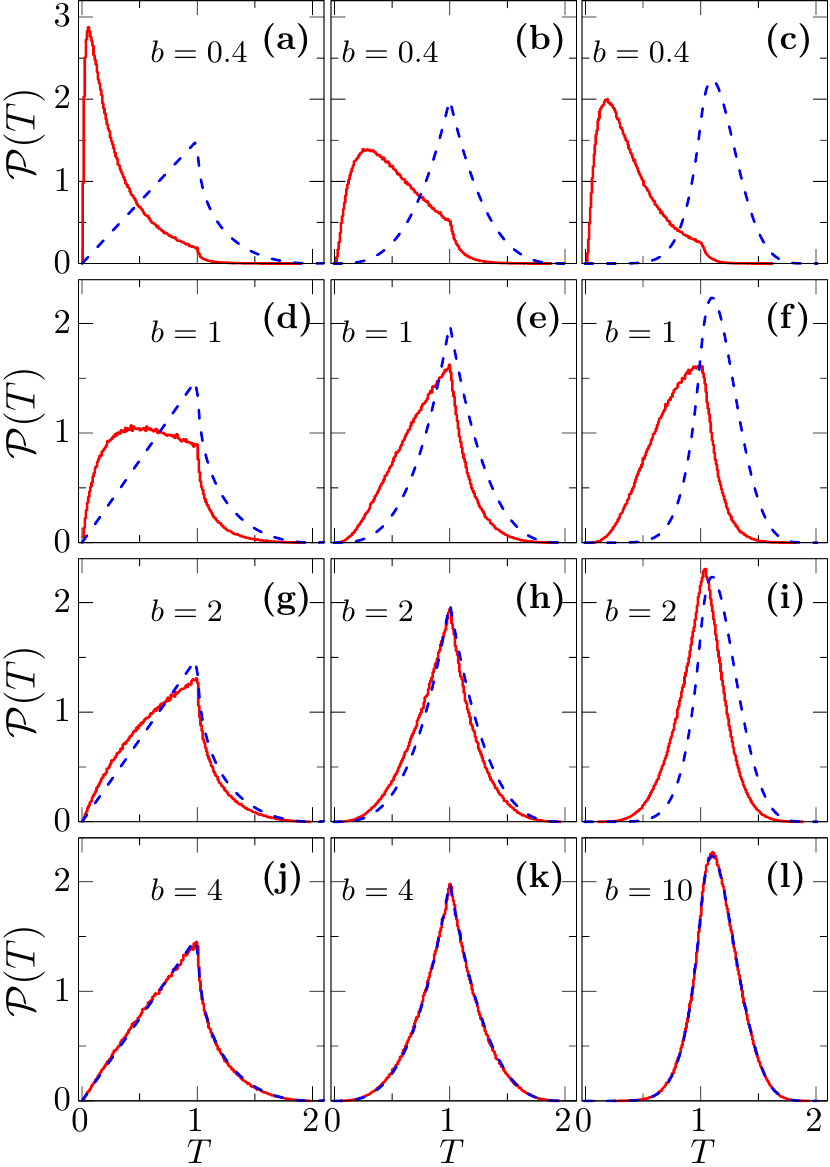}
\caption{Transmission distribution of the PBRM model at criticality for $M=4$ and several values of the effective bandwidth $b$, as indicated in the panels. The symmetries are $\beta=1, 2,$ and 4 in the left, middle, and right columns, respectively. The red lines are histograms obtained by numerical simulations and the dashed blue lines are the corresponding RMT predictions~(\ref{eq:PofTB1M=2})-(\ref{eq:PofTB4M=2}). }
\label{fig:PofTM=2}
\end{figure}

\begin{figure}
\centering
\includegraphics[width=0.48\textwidth]{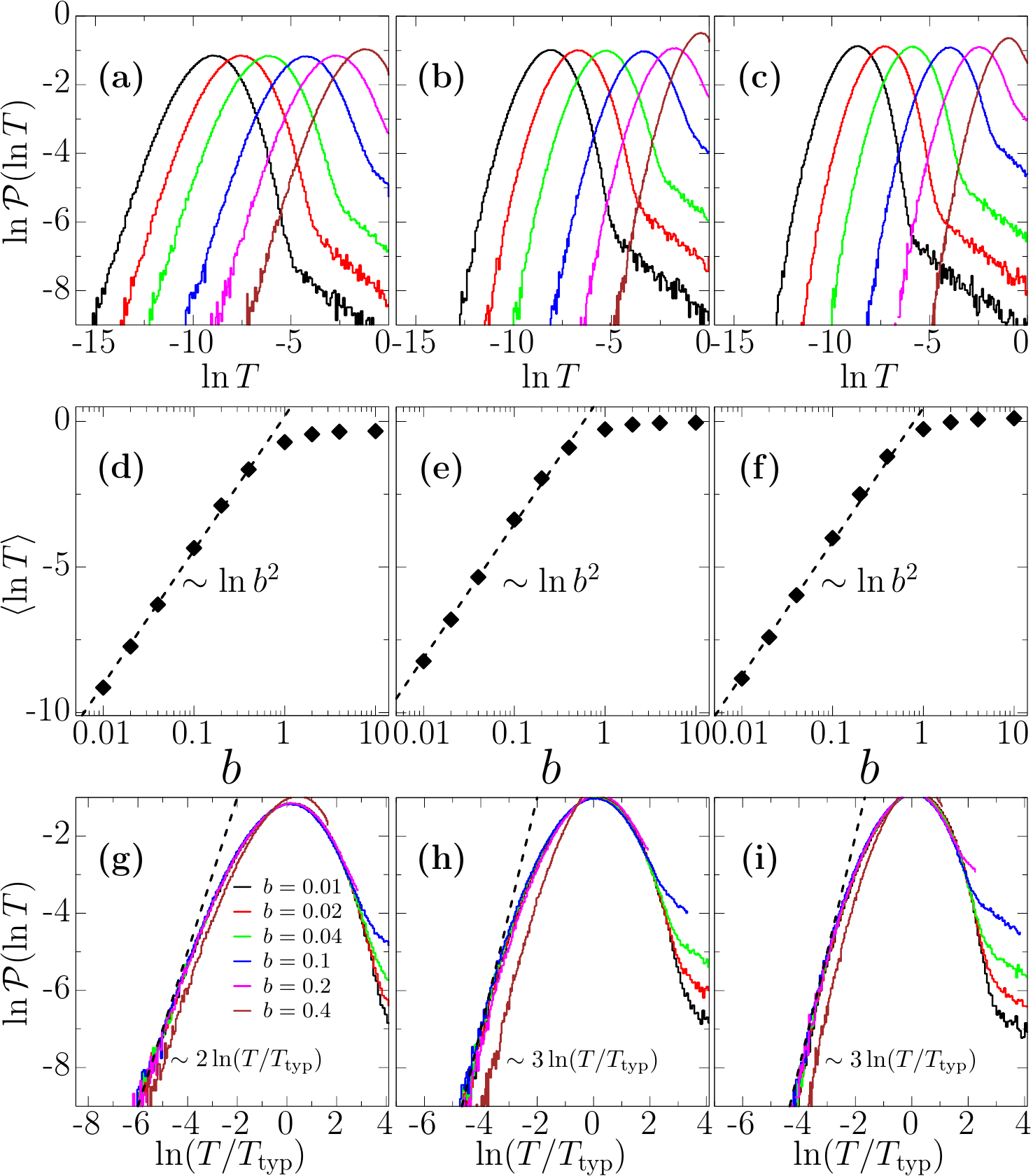}
\caption{Distribution of the logarithm of transmission of the PBRM model at criticality with $M=4$ and several values of the bandwidth $b$, as indicated in panel (g). The symmetries are $\beta=1$ (first column), 2 (middle column), and 4 (third column). The error bars in panels (d)-(f) are smaller than the symbol size. }
\label{fig:lnP_lnT_M=2}
\end{figure}

The transmission distribution for the PBRM model at criticality with four single-channel leads attached it ($M = 4$) is reviewed in Fig.~\ref{fig:PofTM=2}. The symmetries are $\beta=1, 2,$ and 4, in the left, middle, and right panels, respectively, and the considered values of $b$ are indicated in the panels. The red lines are histograms obtained from numerical simulations, and the dashed blue lines are the RMT predictions~(\ref{eq:PofTB1M=2})--(\ref{eq:PofTB4M=2}) according to the symmetry class. For each case, a smooth transition from an insulator-like to a metallic-like regime is observed. In Figs.~\ref{fig:PofTM=2}(j) and 
\ref{fig:PofTM=2}(k), the metallic-like regimen is reached for values of $b=4$ when the system is in the presence of the orthogonal and the unitary symmetry, respectively. However, when the system is in the presence of the symplectic symmetry, a larger value of $b$ (= 10) is required to reach the metallic-like phase, as shown in Fig.~\ref{fig:PofTM=2}(l). Close to the insulator-like regime ($b<0.5$), the transmission distribution shows a pretty similar behavior to that obtained for the $M=2$ case, as observed in Fig.~\ref{fig:lnP_lnT_M=2} (see also Fig.~\ref{fig:lnP_lnT_M=1}). There, $\ln\p(\ln T)$ also shows a scaling property, 
i.e.,~all the curves fall on top of each other when normalized to its typical value $T^{\text{typ}}=\exp\bra\ln T \ket \sim b^{2}$, as observed in 
Figs.~\ref{fig:PofTM=2}(g), \ref{fig:PofTM=2}(h), and \ref{fig:PofTM=2}(i). 
Also, $\bra\ln T\ket$ is a linear function of $\ln b$ for $b<1$, see 
Figs.~\ref{fig:PofTM=2}(d), \ref{fig:PofTM=2}(e), and \ref{fig:PofTM=2}(f).

%

For the PBRM model at criticality with $\beta=1$ symmetry, it has been conjectured that the averages $\bra T\ket,\;\text{Var}(\bra T\ket)$, and $P$, as a function of the bandwidth $b$, obey the following expression
\begin{equation}
\label{eq:cunjectureavgX}
X(b) = X_{\text{RMT}} \left[ \frac{1}{1+(\delta\;b)^{-2}} \right] ,
\end{equation}
where $X$ may be $\bra T\ket,\;\text{Var}(\bra T\ket)$, or $P$; $X_{\text{RMT}}$ is the corresponding RMT prediction and $\delta$ is a fitting parameter. In what follows, this conjecture is verified for the PBRM model at criticality for the $\beta=2$ and 4 symmetries. For completeness, the $\beta=1$ case is also reviewed. For the numerical analysis, the wire length is set to $N=200$ and $10^{5}$ realizations of the $S$ matrix of Eq.~(\ref{eq:scatteringmatrix2}) are generated.

%
\begin{figure}
\centering
\includegraphics[width=0.48\textwidth]{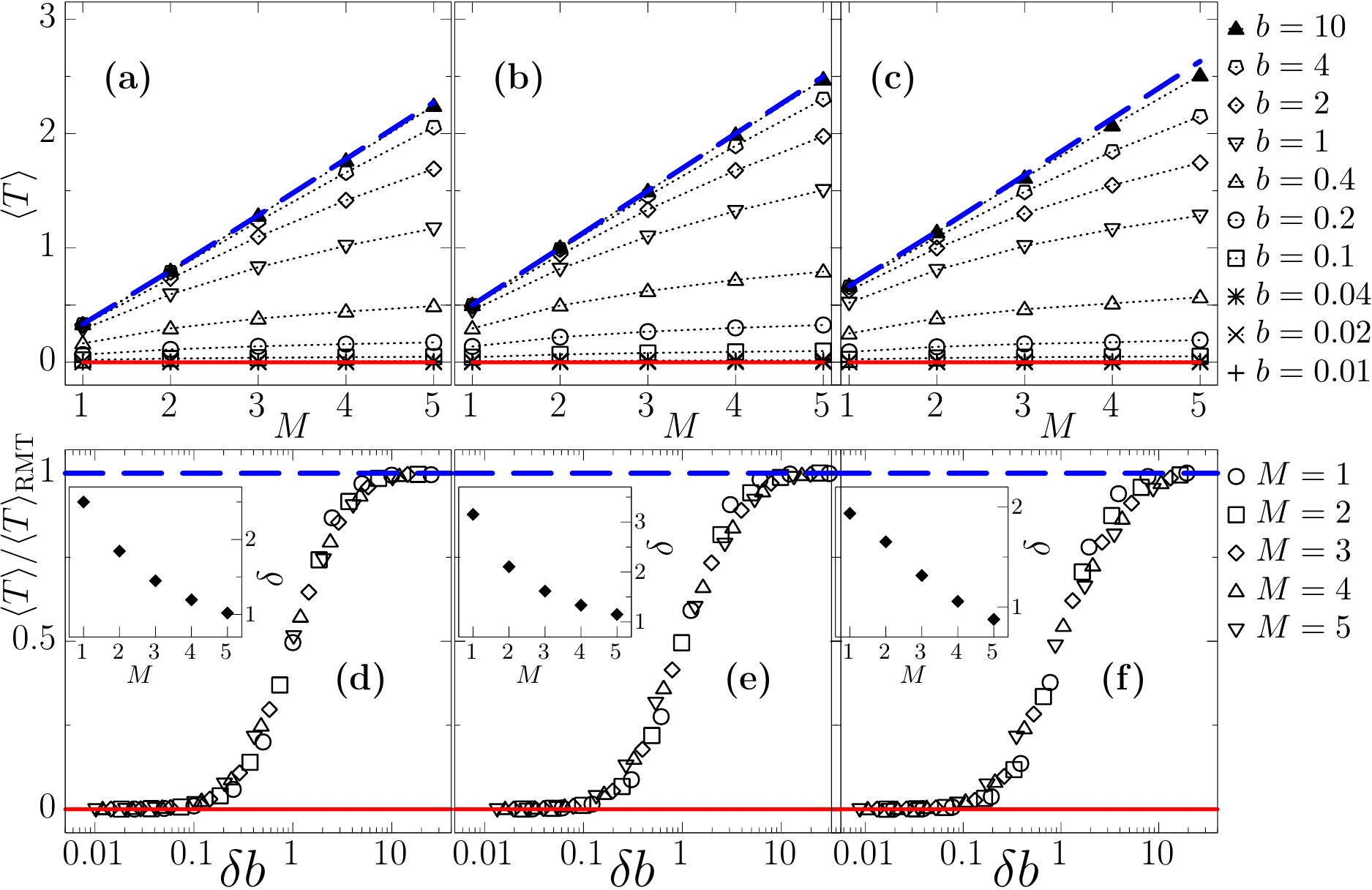}
\caption{Average transmission for the PBRM model at criticality with symmetry $\beta=1, 2, $ and 4, in the first, second, and third columns, respectively. Top panels: $\bra T\ket$ as a function of $M$ for several values of the bandwidth $b$. The symbols correspond to numerical simulations, the dashed blue lines are the RMT predictions~(\ref{eq:avgT}), and the red line at $\bra T\ket=0$ is plotted to guide the eye. Bottom panels: Conjecture~(\ref{eq:cunjectureavgX}) for $\bra T\ket$, the symbols correspond to numerical simulations. Insets: Fitting parameter from~(\ref{eq:cunjectureavgX}) for each $M$, the error bars are the rms of the residuals and are smaller than the symbol size. }
\label{fig:avgT2M}
\end{figure}

In Figs.~\ref{fig:avgT2M}(a), \ref{fig:avgT2M}(b), and \ref{fig:avgT2M}(c), for the PBRM model at criticality $\bra T\ket$ as a function of the open channels $M$ is shown for $\beta=1, 2,$ and $4$, in the first, second, and third columns, respectively, and for several values of $b$. The symbols correspond to numerical results while the dashed blue lines are the RMT predictions~(\ref{eq:avgT}) according to the symmetry class. The red horizontal line at $\bra T\ket=0$ is shown to guide the eye. In those panels, it is observed that for the three symmetry clases at small $b<0.2$, $\bra T\ket\approx0$ since the system is in the insulator-like regime. As $b$ increases ($0.2\leq b<4$) the average transmission also increases until it reaches the RMT prediction ($b=10$) where the system is in the metallic-like regime. For the three symmetry classes labeled by $\beta$, Eq.~(\ref{eq:avgT}) gives an accurate description even though deviations appear for the $\beta=4$ case when $M\ge 4$. In the bottom panels the conjecture~(\ref{eq:cunjectureavgX}) for $\bra T\ket$, normalized to its RMT prediction~(\ref{eq:avgT}), as a function of $\delta b$ for $\beta=1$ [Fig.~\ref{fig:avgT2M}(d)], 
2 [Fig.~\ref{fig:avgT2M}(e)], and 4 [Fig.~\ref{fig:avgT2M}(f)], is shown. The symbols are obtained from numerical simulations while the dashed blue and dashed red lines, at $\bra T\ket=1$ and 0 respectively, are shown to guide the eye. In this case, the conjecture correctly describes the numerical data as can be seen by the shape described by the symbols and more clearly in the small error bars shown in the insets, which are not even visible since they are smaller than the symbols.

%
\begin{figure}
\centering
\includegraphics[width=0.48\textwidth]{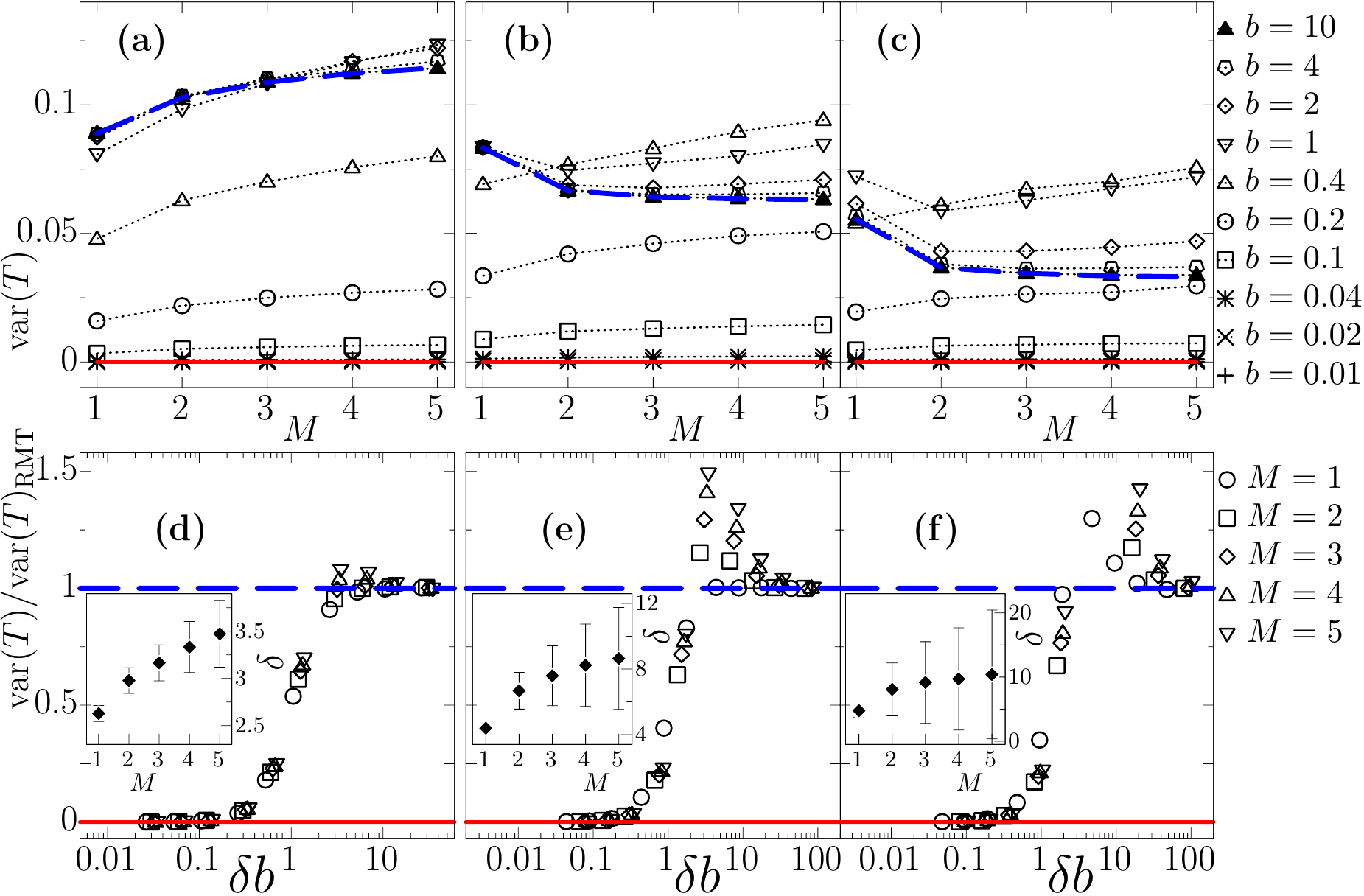}
\caption{Variance of transmission for the PBRM model at criticality in the presence of the symmetry $\beta=1$ (first column), 2 (second column), and 4 (third column). Top panels: $\text{var}(T)$ as a function of $M$ for several values of the bandwidth $b$, the symbols are the numerical results, dashed blue lines are the RMT predictions~(\ref{eq:avgVarT}), and the red line at $\text{var}(T)=0$ is shown to guide the eye. Bottom panels: Conjecture~(\ref{eq:cunjectureavgX}) for $\text{var}(T)$, normalized to its corresponding RMT prediction. The symbols correspond to numerical simulations while the dashed blue line is the expression~(\ref{eq:cunjectureavgX}). Insets: Fitting parameter from~(\ref{eq:cunjectureavgX}) for each $M$. The error bars are the rms of the residuals. }
\label{fig:avgVarT2M}
\end{figure}

In Fig.~\ref{fig:avgVarT2M}, the results for the variance of $T$ as a function of $M$ for the PBRM model at criticality are reported for several values of $b$ and for the symmetries $\beta=1, 2$, and 4, in the first, second, and third columns, respectively. In the top panels, the symbols are the numerical data, the dashed blue lines are the corresponding RMT prediction~(\ref{eq:avgVarT}), and the red line at $\text{var}(T)=0$ is shown to guide the eye. It can be observed that for $M\geq2$ and certain interval of $b$, the fluctuations are greater than the RMT predictions. That behavior is more notorious for the $\beta=2$ and 4 cases in the interval $0.4\leq b\leq 4$. However, in the metallic regime ($b=10$), a good agreement with the RMT predictions is obtained. 

The $\text{var}(T)$ normalized to its corresponding RMT prediction as a function of $\delta b$ is plotted in Fig.~\ref{fig:avgVarT2M}(d), \ref{fig:avgVarT2M}(e), and \ref{fig:avgVarT2M}(f), for $\beta=1, 2,$ and 4, respectively. For $2<\delta b<40$ and $M\geq2$ large deviations between the conjecture~(\ref{eq:cunjectureavgX}) (dashed blue line) and the numerical data (symbols) are observed, even though these deviations are smaller for the $\beta=1$ case. This can also be appreciated in the insets where the error bars are relatively large.

\begin{figure}
\centering
\includegraphics[width=0.48\textwidth]{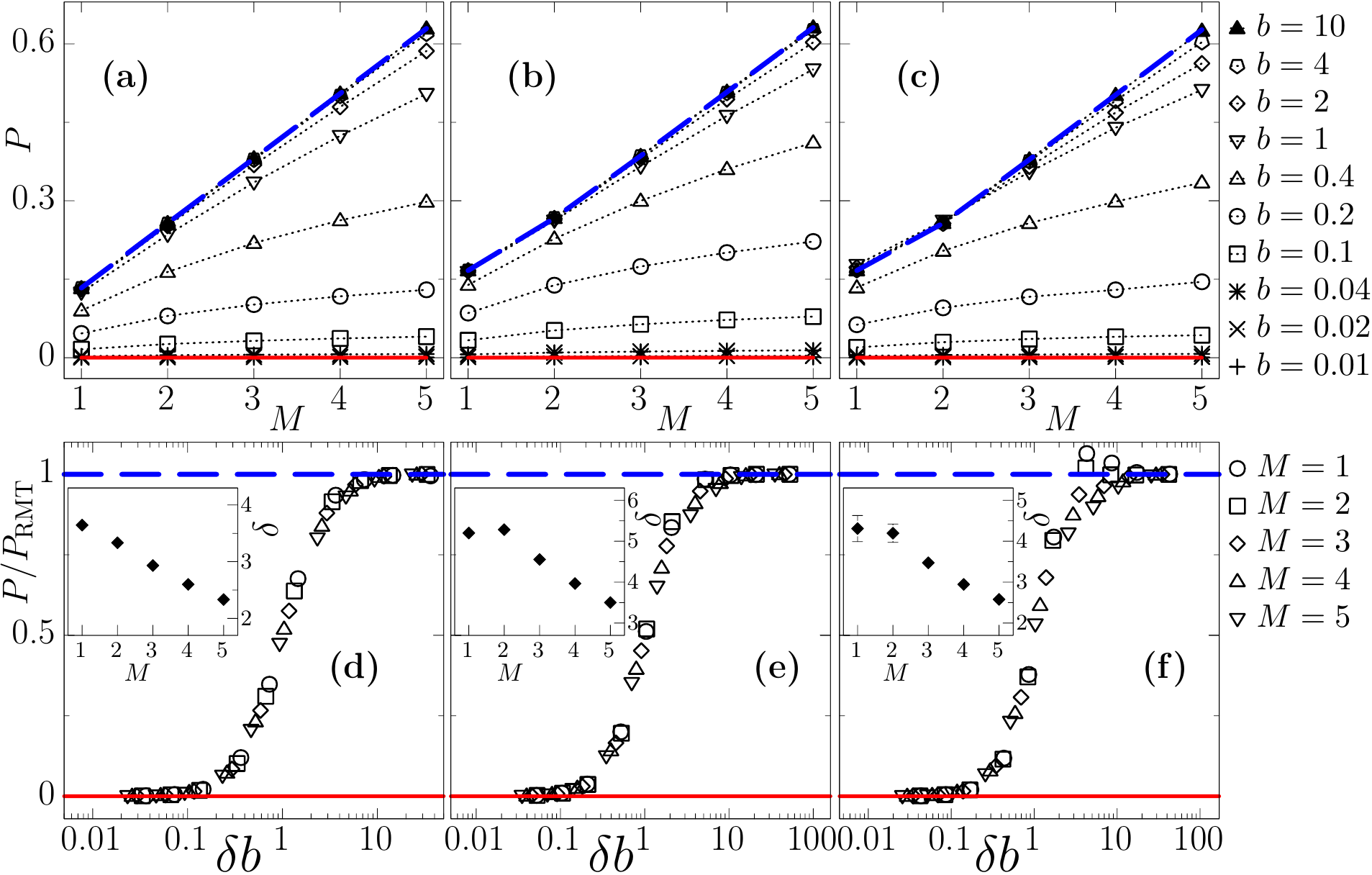}
\caption{Shot noise power $P$ for the PBRM model at criticality with symmetry $\beta=1$ (first column), 2 (second column), and 4 (third column). Top panels: $P$ as a function of $M$ for several values of the bandwidth $b$, the symbols correspond to numerical results, the dashed blue lines are the RMT predictions~(\ref{eq:shotnoise}), and the red line at $P=0$ is shown to guide the eye. Bottom panels: Conjecture~(\ref{eq:cunjectureavgX}) for $P$, the symbols are obtained by numerical simulation. Insets: Fitting parameter from~(\ref{eq:cunjectureavgX}) for each $M$, the error bars are the rms of the residuals. }
\label{fig:shotnoiseP}
\end{figure}

Finally, the shot noise power $P$ as a function of $M$ and for several values of $b$ is reported in the top panels of Fig.~\ref{fig:shotnoiseP}. A transition from an insulator-like ($b=0.01$) where $P=0$ (red horizontal line) to a metallic-like ($b=10$) regime (symbols) is clearly observed. In the metallic-like regime, the numerical results (symbols) are in agreement with the RMT predictions~(\ref{eq:shotnoise}) (dashed blue lines). In the bottom panels of the same figure, the conjecture~(\ref{eq:cunjectureavgX}) for $P$ as a function of $\delta b$ and several values of $M$ is shown. A good agreement between the numerical data and expression~(\ref{eq:cunjectureavgX}) is obtained for the three symmetry classes $\beta$, as revealed by the shape described by the symbols and the small error bars obtained (see insets). 

%
%


\section{Conclusions}
\label{sec:Conclusions}

In this paper, an extensive numerical study of the scattering and transport properties of the PBRM model at criticality in the presence of the three symmetry classes, orthogonal, unitary, and symplectic, has been presented. For the sake of completeness, some known results for the orthogonal case previously studied in the literature have also been reviewed, while new ones for the unitary (in the periodic model) and the symplectic cases, which had remained unexplored in the context of the PBRM models, are reported. Surprisingly, the results presented confirm that the scattering and transport properties of the symplectic PBRM model can be well described by existing analytical and heuristic relations widely used in studies of the PBRM model in the presence of the $\beta=1$ and 2 symmetries. Importantly, for the three symmetry classes the multifractal properties of the isolated model were obtained from scattering and transport properties, which is very convenient from the experimental point of view since direct access to the eigenfunctions is not required. 

Additionally, an analytical result for the transmission distribution in the presence of the symplectic symmetry with $M=4$ open channels~(\ref{eq:PofTB4M=2}), which applies to the symplectic PBRM model at criticality in the metallic regime, was provided. Also, in this study, the results for $\beta=1$ are in accordance with those reported in, for example, Ref.~\cite{Mendez2010}. Moreover, to our knowledge, our results for $\beta=2$ have not been reported before; only the nonperiodic version of the PBRM model with broken time-reversal symmetry ($\beta=2$) has been studied in Ref.~\cite{Alcazar2009}. And the symplectic case, $\beta=4$, has not been reported in neither the periodic nor in the nonperiodic version of the PBRM model so far. Thus, with the present study a more clear panorama about the scattering and transport properties of the PBRM model at criticality in the presence of the three classical Wigner-Dyson symmetries is given.

\vfill


\vspace{0.1cm}

\acknowledgments

We thank I. Varga for his useful comments on the manuscript. A. M. M-A acknowledges financial support from CONACyT. 
M.C.-N. thanks financial support from CONACyT (Grant ``Ciencia de Frontera 2019'', No. 10872) and the facilities 
provided by the ``Centro de An\'alisis de Datos y Superc\'omputo (CADS)'' from the University of Guadalajara 
through of the Leo-Atrox Supercomputer. 
J.A.M.-B. thanks support from CONACyT (Grant No. 286633), CONACyT-Fronteras (Grant No. 425854), 
VIEP-BUAP (Grant No. 100405811-VIEP2022), and Laboratorio Nacional de Superc\'omputo del Sureste 
de M\'exico (Grant No. 202201007C), Mexico.





\end{document}